# Mars Express measurements of surface albedo changes over 2004 - 2010

*M. Vincendon[1]; J. Audouard[1]; F. Altieri[2]; A. Ody[1,3].*

[1]*Institut d'Astrophysique Spatiale, Université Paris Sud, 91405 Orsay, France (mathieu.vincendon@ias.u-psud.fr, phone +33 1 69 85 86 27, fax +33 1 69 85 86 75).*

[2]*INAF, IAPS, Rome, Italy.*

[3]*Laboratoire de Géologie de Lyon, 69622 Villeurbanne, France.*

**Abstract**

The pervasive Mars dust is continually transported between the surface and the atmosphere. When on the surface, dust increases the albedo of darker underlying rocks and regolith, which modifies climate energy balance and must be quantified. Remote observation of surface albedo absolute value and albedo change is however complicated by dust itself when lifted in the atmosphere. Here we present a method to calculate and map the bolometric solar hemispherical albedo of the Martian surface using the 2004 - 2010 OMEGA imaging spectrometer dataset. This method takes into account aerosols radiative transfer, surface photometry, and instrumental issues such as registration differences between visible and near-IR detectors. Resulting albedos are on average 17% higher than previous estimates for bright surfaces while similar for dark surfaces. We observed that surface albedo changes occur mostly during the storm season due to isolated events. The main variations are observed during the 2007 global dust storm and during the following year. A wide variety of change timings are detected such as dust deposited and then cleaned over a Martian year, areas modified only during successive global dust storms, and perennial changes over decades. Both similarities and differences with previous global dust storms are observed. While an optically thin layer of bright dust is involved in most changes, this coating turns out to be sufficient to mask underlying mineralogical near-IR spectral signatures. Overall, changes result from apparently erratic events; however, a cyclic evolution emerges for some (but not all) areas over long timescales.

# 1 Introduction

The changing appearance of Mars to human eyes has long been documented from Earth and satellite observations. Major changes are related to atmospheric and polar phenomena occurring on daily to seasonal scales: formation, movement and dissipation of ice and dust clouds; advance and retreat of polar caps (Leighton, et al., 1969; Baum, 1974; Pleskot & Miner, 1981). Changes also occur at the ice-free surface on multiple timescales. These variations are essentially due to modification of the surface coating by fine bright dust over the larger-grained darker underlying material (Pollack & Sagan, 1967; Sagan, et al., 1973). Episodic removal of large amount of bright dust due to short-lived strong winds occurs during the storm season (De Mottoni Y Palacios & Dollfus, 1982; Lee, 1986; Szwast, et al., 2006; Cantor, 2007; Geissler, et al., 2010). Progressive dust erosion is also observed under the effect of (1) topographically and thermally driven horizontal wind gusts (Sagan, et al., 1973; Christensen, 1988; Szwast, et al., 2006; Cantor, et al., 2006), (2) dust devils (Geissler, 2005; Greeley, et al., 2005; Greeley, et al., 2006; Cantor, et al., 2006) and (3) dark sand saltation (Sullivan, et al., 2008; Geissler, et al., 2010; Vaughan, et al., 2010). Subsequent settling of dust (Landis & Jenkins, 2000) then brightens dark surfaces. Local (Geissler, et al., 2010) or regional (Szwast, et al., 2006) horizontal transport of dust by episodic strong winds is observed during the storm season, as well as gradual accumulation of dust linked with topography (Veverka, et al., 1977; Geissler, 2012). Regional fallout also temporarily blankets certain areas of Mars such as Syrtis Major during global dust storm decay (Christensen, 1988; Szwast, et al., 2006; Cantor, 2007). Other minor contributors to surface albedo change include horizontal movements of dark sand grains under strong winds (Geissler, et al., 2010; Bridges, et al., 2012; Chojnacki, et al., 2014), gravity-driven mass-movements (Sullivan, et al., 2001; Treiman, 2003; Schorghofer & King, 2011), defrosting processes modifying underling ice-free surface (Dundas, et al., 2014), and slope flow potentially involving water (Malin, et al., 2006; McEwen, et al., 2011).

The existence and degree of persistence of albedo modification depends on the competition between erosion and accumulation mechanisms, which relative intensity depends on season, year, location, and surface properties (De Mottoni Y Palacios & Dollfus, 1982; Lee, 1986; Szwast, et al., 2006; Sullivan, et al., 2008; Geissler, et al., 2010). Overall, dark regions seem to rapidly recover their pre-storm albedo value (Pleskot & Miner, 1981; Christensen, 1988; Cantor, 2007; Vincendon, et al., 2009), suggestive of a relatively perennial albedo distribution, in agreement with the persistence of Mars main albedo markings over time (Christensen, 1988; Geissler, 2005; Szwast, et al., 2006). However, significant local long-term differences, lasting over decades, have also been reported (Pollack & Sagan, 1967; Baum, 1974; De Mottoni Y Palacios & Dollfus, 1982; Capen, 1976; James, et al., 1996; Bell, et al., 1999; Erard, 2000; Geissler, 2005), as well as progressive movements of albedo frontiers (Veverka, et al., 1977; Chaikin, et al., 1981; Geissler, 2005). Assessing the irreversible versus cyclic nature of these changes requires further albedo change monitoring.

Surface albedo changes and associated dust reservoir redistribution could alter current climate via temperatures and winds modification (Kahre, et al., 2005; Cantor, 2007; Fenton, et al., 2007; Montmessin, et al., 2007). Albedo is indeed a key parameter controlling the energy budget at the surface and must be properly modeled in energy balance codes and global climate models (Forget, et al., 1999; Kieffer, 2013), as well as during thermal inertia retrievals (Mellon, et al., 2000; Putzig, et al., 2005; Fergason, et al., 2006; Audouard, et al., 2014). Variations in dust coating also modify the distribution and apparent spectral properties of exposed surface material detected by remote sensors (Singer & Roush, 1983; Christensen, 1988; Poulet, et al., 2007; Rice, et al., 2011; Carrozzo, et al., 2012). Understanding the timing and mechanisms associated with current dust deposition and removal is also of importance for future mission planning, as persistent settling of dust alters robotic and instrumental performances (Landis & Jenkins, 2000; Smith, et al., 2006; Kinch, et al., 2007; Vaughan, et al., 2010; Drube, et al., 2010; Lemmon, et al., 2015).

Estimating absolute surface reflectance values and associated time variations from orbit is tricky. First, remote observations of the surface are performed with variable illumination and viewing geometries depending on spacecraft, season and latitude, which results in phase effects of both the atmosphere and the surface. Second, observations of the surface are performed through the atmosphere which contains varying amounts of dust and clouds. Third, the reflectance at all solar wavelengths and in all directions must be accounted for to get precise constraints about the energy balance at the surface. Finally, the time sampling of observations is also of importance to assess the duration of detected changes. The OMEGA imaging spectrometer onboard Mars Express measures wavelengths from early visible to near-IR. Global coverage of the planet with repeated observations of several areas with variable photometric and atmospheric conditions have been obtained over more than 3 Mars Years (MY), from early 2004 (late MY26) to mid-2010 (mid MY30). Additional observations with a restricted wavelength range are ongoing (2014, MY32). This time range includes the 2007 (MY28) global dust storm (GDS), whose timing significantly differs from the 2001 (MY25) storm monitored by Mars Global Surveyor (MGS). The spatial resolution of OMEGA varies between 0.3 and 5 km, which provides a compromise between global coverage and details at the surface. In this study, we calculate surface hemispherical bolometric albedo using OMEGA data and report surface changes observed over the last 10 years at non polar latitudes (60°S – 60°N).

## 2  Method

### 2.1  OMEGA data characteristics and selection

The OMEGA dataset consist of hyperspectral images of the surface and atmosphere of Mars obtained from orbit. OMEGA collects the reflected sunlight between 0.35 and 5.1 µm over 352 spectral channels divided into three detectors, with a contribution of thermal emission for wavelengths greater than 3 µm. Ground photometric calibration was performed for a targeted accuracy better than 20% in absolute terms (Bonello, et al., 2005). Comparison of OMEGA data with HRSC and telescopic observations in the visible and near-IR during the mission showed that the absolute accuracy is in fact better than 10% in that range (McCord, et al., 2007). The calibration of the 0.36 – 1.07 µm channel (hereafter "visible channel") has been recently revisited, with a high level of confidence for wavelengths ≥ 0.43 µm and ≤ 0.95 µm (Bellucci, et al., 2006; Carrozzo, et al., 2012). The photometric response of the 0.93 – 2.69 µm channel (hereafter "near-IR channel") is checked for each orbit via an on-board calibration procedure and remained stable during the mission (Ody, et al., 2012). We will only consider wavelengths ≥ 1.08 µm for this channel to retain data with the highest signal to noise ratio. While atmospheric gas absorptions can be corrected for wavelengths ≤ 2.5 µm (see e.g. (Langevin, et al., 2007)), a changing water vapor feature followed by a broad saturated $CO_2$ gas band makes it difficult to recover surface reflectance beyond 2.5 µm. The photometric response of the 2.53 – 5.09 µm channel had varied over the mission, making use of this channel for absolute reflectance measurements more complex (Jouglet, et al., 2009). This channel also contains numerous broad, partly saturated gas absorptions (for wavelengths ≤ 2.9 µm and ≥ 4 µm), and covers a wavelength range for which the solar flux is negligible (see 2.2) while the thermal emission from the surface is significant: it will thus not be used in this work.

The spatial resolution at the surface ranges from 0.3 km to 5 km depending on Mars Express altitude on its elliptical polar orbit. The spatial dimensions of images correspond to a few hundreds to thousands of pixels in the orbit direction and to 16 – 128 pixels perpendicularly to the orbit (the lower the spacecraft altitude is, the smaller the image width is). Two co-aligned units cover the solar range with distinct telescopes and operating modes: the push broom method for the visible channel (with a line of pixels perpendicular to the orbit) and the whisk broom technology for the near-IR channel (with one pixel scanning the surface perpendicularly to the orbit). These two acquisition methods and the fact that the visible channel is adjacent but physically separated from the near-IR one produce some differences in the spatial coverage of the two channels. First, the field of view is larger for the visible channel compared to the near-IR channel, especially in the "Y" direction (i.e., along the spacecraft track), which partly lowers the actual spatial resolution when we combine both detectors. Second, a misalignment between these two channels results in spatial registration inconsistency of typically a few pixels. This misalignment changes from observation to observation (Carrozzo, et al., 2012) and within a given observation. We co-register both channels by maximizing the correlation between the 0.9 µm and 1.1 µm measurements with a changing shift value within a given observation. In the "X" direction (i.e., perpendicular to the spacecraft track) the misalignment ranges from 2 pixels (for 16 pixels wide observations) to 5 – 6 pixels (for 128 pixels wide observations). In the Y direction the shift is greater than 3 pixels. It depends on the spacecraft altitude and velocity which vary as Mars Express orbit is elliptical. Thus, the Y misalignment can change by several pixels within a given observation. This procedure can however not be easily achieved automatically for cubes including targets other than the surface (typically, cubes observing both the surface and the atmosphere at the limb). These observations will thus be considered but as lower quality data.

Viewing conditions vary within the dataset. Emission (emergence) angles are near-nadir for most observations but can be up to 90° when pointing drifts toward limb observations. Solar zenith angles (also referred as incidence angles) ranges from zenith (0°) to terminator (90°) depending on local time, season and latitude. Due to these variations in viewing geometries, OMEGA is particularly sensitive to atmospheric and surface phase effects. Our analysis will be primarily based on near-nadir data (emergences lower than 10°) obtained with incidences lower than 75° to limit shadowing and strong phase effects. Observations with higher photometric angles will also be used as lower quality data.

Various instrumental caveats have to be accounted for while processing the global OMEGA dataset in terms of absolute reflectance values (Langevin, 2007; Jouglet, 2008; Ody, et al., 2012). Some data must be removed: calibration lines at the beginning of certain cubes, as detailed in the "README" files of the omega software pipeline; the four last lines of each cubes, frequently corrupted; cubes tagged with a 0 in their data_quality header keyword; heating cubes. Other issues do not necessarily weaken the ability of OMEGA to provide an accurate measure of the solar bolometric radiance (Table 1) but must be handled carefully:

- Near-IR channel detector performances have been designed and calibrated for a detector temperature lower than 80K. Such a temperature has been obtained during almost the entire mission, except for some MY30 data as cryocoolers performance started to decline at that time. These higher temperatures essentially result in a higher amount of noise and a greater dark current level. This weakens OMEGA capability to detect narrow spectral features and reduce the detector dynamic, but does not significantly change the value of the integrated radiance over a broad wavelength range: the temperature threshold can then be increased to -177°C instead of -194°C to recover more data.
- A given observation can cover a wide range of latitude and thus terrain types and lighting conditions. Saturation can then occur when observed surface was brighter than expected during exposure time calculation due the presence of ice or sun exposed slopes. OMEGA near-IR detector uses a "pre-charge" design: photons reduce charges from about 4000 digital numbers (hereafter "DN") and saturation occurs when a level of about 330 DN is reached. Saturation of the near-IR channel first occurs at wavelengths about 1.5 µm where the balance between near-IR detector sensitivity and solar flux is the most favorable. While approaching detector limit, non-linearity effects are observed and can result in false absorption bands about 1.5 µm. A conservative threshold of 500 DN is then considered for spectel #40 ($\lambda$ = 1.486 µm) while looking for features in that range. However, integrated flux measurements are not strongly sensitive to these non-linearity issues restricted to a narrow wavelength range and data down to a threshold of 340 DN can be used. Visible channel saturation occurs above 4095 DN (digital saturation) / 4040 DN (physical saturation). A threshold of 4020 DN has been selected and applied to spectel #300 (0.6848 µm) to filter data.
- A particular instrumental issue perturbs 128 pixels wide observations gathered between orbits 513 and 2123. For these cubes, columns #81 – 96 (12.5% of columns) are corrupted over a set of 44 wavelengths regularly spaced over the whole wavelength range, with two different sets alternating one line out of two. Corrupted wavelengths can be recovered using spatial or spectral interpolation: this second solution will be used here. Between orbits 2124 and 3283, the same issue occurred, plus a more random noise that affects half of the cube (columns #65 – 128). This second noise cannot be corrected and the corrupted part of the cubes must be removed.

We have defined three quality data levels which are detailed in Table 1. High quality data are obtained by applying the strongest restriction as mentioned above, while lower quality data include data with issues or lower viewing conditions. These lower quality data will be considered to fill gaps or increase time sampling.

Table 1 : Quality levels of OMEGA global maps. Observations conditions and instrumental issue criteria thresholds are indicated for the 3 quality levels. Level #1 is higher quality data; level #2 includes data with instrumental issues of low impact on bolometric albedo; level #3 includes data with lower photometric/atmospheric conditions. Incidence and emergence refers to angles measured from the outward normal to the reference ellipsoid. Water ice corresponds to the 1.5 µm band depth as described in (Ody, et al., 2012). Saturation threshold is applied to spectel #40 ($\lambda$ = 1.486 µm).

| quality level | incidence | emergence | water ice threshold | Detector temperature | Saturation threshold (near-IR) | Corrupted 128 pixel cubes recovered | Non coregistred cubes included |
|---|---|---|---|---|---|---|---|
| 1 (high) | 75° | 10° | 4% | -194°C | 500 DN | no | no |
| 2 | 80° | 20° | 7% | -177°C | 340 DN | yes | no |
| 3 (low) | 80° | 30° | 10% | -177°C | 340 DN | yes | yes |

## 2.2 Integrating OMEGA wavelengths to derive bolometric albedo

Solar albedo is the ratio between the emitted and received flux over the whole solar spectral range. OMEGA makes it possible to measure with a great level of confidence surface reflectance between 0.43 µm and 2.5 µm (see previous sections). This range corresponds to 85% of the solar flux (Fröhlich & Lean, 2004). Another 11.5% of the solar flux is contained in the [0.25 µm – 0.43 µm] UV range. HST observations (Bell & Ansty, 2007) show that the reflectance of Mars for wavelengths lower than 0.43 µm is dominated by the ubiquitous ferric absorption: the reflectance is low (< 0.05) with little spectral and spatial variations. We extrapolated OMEGA reflectance at 0.43 µm down to 0.25 µm using typical surface spectra from (Bell & Ansty, 2007) scaled to match OMEGA measurements at 0.43 µm. The solar flux is negligible below 0.25 µm (< 0.2%). Beyond 2.5 µm, atmospheric absorptions prevail up to 2.9 µm (see section 2.1). Previous Mars solar albedo measurements typically stop at 2.9 µm (Christensen, et al., 2001). We also neglect here the remaining 2% of the solar flux at longer wavelengths due to instrumental constraints (see 2.1). This approach makes it possible to obtain a spectrum of surface hemispherical reflectance over 96.5% of the solar flux. This spectrum is then summed over the solar spectrum, and divided by the solar flux to derive bolometric albedo.

OMEGA wavelength sampling has varied over the mission due to the progressive aging of the detector (37 spectels over 352 were dead or significantly corrupted in mid-2010) and due to sporadic instrumental problems (128 pixel wide cubes acquired between Mars Express orbits 513 and 3283 are corrupted over specific wavelengths, see 2.1). For each cube, a list of usable wavelengths is defined, from which 70 regularly spaced wavelengths are extracted (40 in the visible channel and 30 in the near-IR channel). The bolometric albedo calculation procedure, in particular the aerosols correction, is then applied on these 70 spectels only to reduce computing time. We have checked that results were not significantly modified by using this limited subset of wavelengths instead of the full spectral range. We have also checked that a given observation leads to similar albedo values whether it is processed with the list of good spectels of the beginning or of the end of the mission. Overall, observed differences between albedos produced by various wavelength sampling and interpolation procedures are lower than 1%.

## 2.3 Aerosols correction

Observations of the surface of Mars by remote sensor are disturbed by scattering and absorption of light within aerosols (Pleskot & Miner, 1981; Lee & Clancy, 1990). It has e.g. been demonstrated that reflectance variations of dark terrains with viewing angles or time are essentially due to aerosols, while surface photometric effects are second order contributors (Pleskot & Miner, 1981; Vincendon, et al., 2009; Vincendon, 2013). The main component of Martian aerosols is suspended dust for which composition is relatively homogeneous due to perpetual lifting, transport and mixing at global scale (Yen, et al., 2005; Goetz, et al., 2005). Mars dust single scattering albedo, phase function and cross section have been recently reevaluated (Tomasko, et al., 1999; Vincendon, et al., 2007; Wolff, et al., 2009). At visible and near-IR wavelengths, dust is bright and forward scattering due to its small micrometer size, with a broad ferric absorption in the UV/visible. The mean particle size of aerosols slightly varies with time and place, and with viewing geometries due to the vertical gradient of particle size; most variations of the effective radius are between 1.2 and 1.8 µm (Chassefiere, et al., 1995; Wolff & Clancy, 2003; Vincendon, et al., 2009; Wolff, et al., 2009; Lemmon, et al., 2015). The amount of dust in the atmosphere can be divided into a background component, for which seasonal variations are relatively homogeneous with longitude and latitude once elevation/pressure is accounted for, and local to regional events particularly frequent during the storm season (L$_S$ 130° - 340°, see e.g. Figure 1) and near the edge of the seasonal caps in local fall and winter (Liu, et al., 2003; Smith, 2004; Lemmon, et al., 2004; Vincendon, et al., 2008; Vincendon, et al., 2009). Water ice aerosols are mainly observed between the equator and 20°N (in particular east of Tharsis) for solar longitude ranging from 60° to 140° (Aphelion cloud belt), above Hellas, and near/above seasonal caps (Smith, 2004).

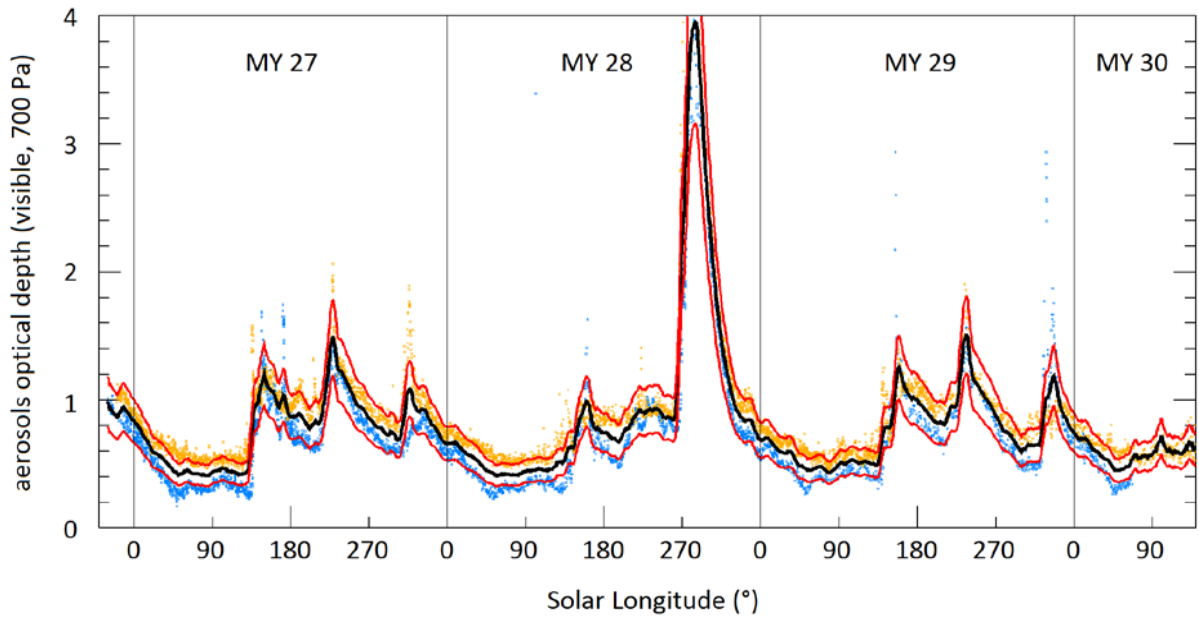

Figure 1 : Visible aerosols optical depth during OMEGA near-IR channel operating period (2004 – 2010), adapted from (Lemmon, et al., 2015). Pancam 0.9 μm measurements at Spirit (blue) and Opportunity (orange) landing sites (Lemmon, et al., 2015) are scaled to a 700 Pa pressure level using a pressure predictor (Forget, et al., 2007). These measurements are averaged and smoothed (black line) to derive the typical time variations of the optical depth at low to mid-latitudes used during data processing (see text). The ± 20% range used for uncertainty estimate is indicated with red lines.

To minimize effect of atmospheric aerosols, the easiest approach is to use observations of the surface of Mars obtained during clear atmospheric conditions with low solar zenith angles (Christensen, et al., 2001). However, even under the clearest atmospheric conditions at least 30% of incoming solar photons interact with aerosols before reaching the surface, which result in substantial effect of atmospheric dust on surface observations (Lee & Clancy, 1990). Another similar approach consists in selecting the darkest observations of a given place (Geissler, 2005). However, if the presence of suspended dust indeed brighten dark surface observed from a nadir remote instrument, it reduces bright surface reflectance observed at nadir as radiative transfer through optically thin aerosols favor high emergences compared to near Lambert surfaces (Lee & Clancy, 1990). Szwast et al. (2006) perform a combined follow-up of solar albedo and thermal-IR dust optical depth to account for dust aerosols while studying the surface. This method may present biases as the solar zenith angle, and hence the aerosols contribution via effective optical depth, change with season and latitude for a given vertical optical depth. Moreover, thermal-IR measurements of the optical depth must be scaled to get visible optical depth, with a scaling factor that is observed to change between 1.5 and 2.5 as a function of particle size (Lemmon, et al., 2004; Smith, et al., 2006; Lemmon, et al., 2015). A proper aerosols correction requires radiative transfer modeling with robust constraints on the visible and near-IR properties of aerosols (Lee & Clancy, 1990; Paige & Keegan, 1994; Christensen, et al., 2001).

Our methodology to perform a first order removal of the contribution of aerosols in our dataset is as follow:

- Observations containing water ice near-IR signature are removed using the 1.5 µm water ice band calculated with the formulae detailed in (Ody, et al., 2012). As water ice clouds can show up at 1.5 µm while being thin with a low impact on observed reflectance (Madeleine, et al., 2012), a slackened threshold of 4% to 10% can be used while considering bolometric albedo calculation (Table 1).
- The contribution of dust is then accounted for through modeling: we use the multiple scattering radiative transfer code of (Vincendon, et al., 2007) to construct look-up tables of surface reflectance as a function of observed reflectance, dust aerosols optical depth, and photometric angles.
- We use the single scattering properties of aerosols detailed in (Wolff, et al., 2009), with some adjustments: the unrealistic backscattering peak produced by the T-matrix calculation (Wolff, et al., 2010) is empirically removed and we consider two effective radius (1.2 µm and 1.8 µm) that frame most size variations. The resulting set of aerosols parameters (Figure 2) has been checked with success against OMEGA data, using similar procedures and data as (Vincendon, et al., 2007; Vincendon, et al., 2009) and extended to visible wavelengths.
- Two methods are used to estimate the optical depth. First, we use the background trend in aerosols optical depth during the Mars Express mission derived from MER/Pancam measurements (Lemmon, et al., 2015) at low latitudes (Figure 1), which provide a good proxy for mid-latitudes (Vincendon, et al., 2009). This makes it possible to have nearly simultaneous measurements at the appropriate wavelength, but does not include regional events. Secondly, we use a preliminary version of the daily optical depth maps of (Montabone, et al., 2015) produced by the merging and interpolating of three thermal–IR datasets that cover OMEGA observing years. This makes it possible to include regional events, but require the use of a constant 2.6 scaling factor between IR and visible wavelengths, while variations of this factor by ± 30% are common throughout the year (Lemmon, et al., 2015). Moreover, these optical depths are derived with different thermal-IR instruments and inconsistencies between retrievals have been noted (Montabone, et al., 2015). These optical depths provided at a constant reference pressure level are then scaled to the local pressure using a pressure predictor (Forget, et al., 2007) coupled to a 32 ppd MOLA altimetry map (Smith, et al., 1999).
- Finally, we estimate surface reflectance and associated uncertainty for a given observation by calculating the mean and standard deviation of 4 reflectance retrievals corresponding to the two particle size (1.2 and 1.8 µm) and to a range of ± 20% in optical depth.

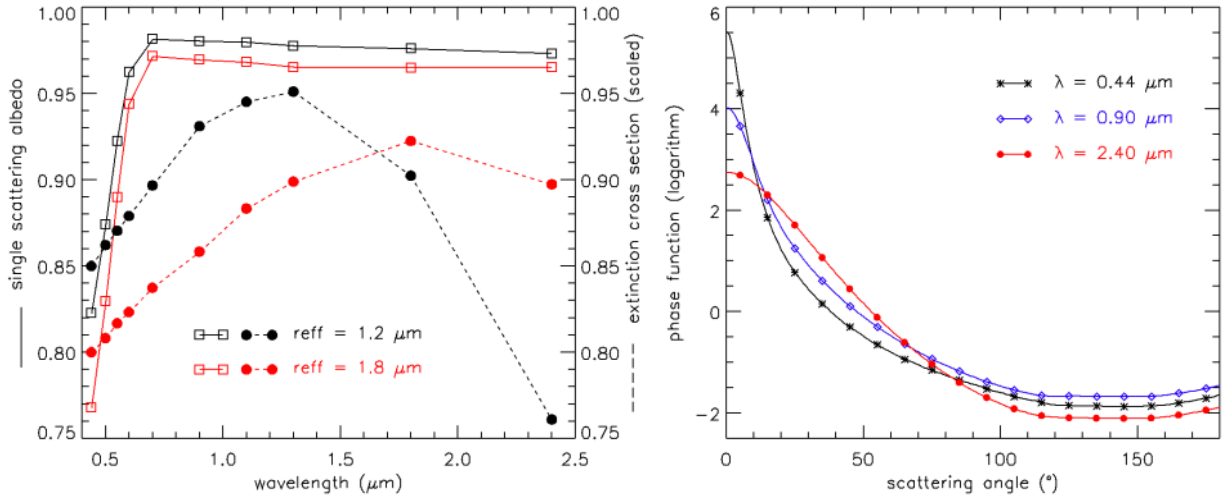

Figure 2: Dust single scattering optical properties used for aerosols correction. Parameters have been calculated using refractive indices and T-matrix parameters published in Wolff et al. (2009) for two effective radii (black, 1.2 µm and red, 1.8 µm; M. Wolff, personal communication), except for the backscattering peak of phase functions (angles > 125°) which is replaced by the Tomasko et al. (1999) phase function shape. (Left) Single scattering albedo (squares) and extinction cross section (dots). (Right) Single scattering phase functions at 3 wavelengths (0.44, 0.9 and 2.4 µm, black stars, blue diamonds and red dots respectively) for an effective radius of 1.2 µm.

## 2.4 Surface phase function

Hemispherical reflectance is the balance, for a given wavelength, between incoming solar radiations and scattered radiations in all direction (Hapke, 1993). This is the useful quantity for energy balance calculation and climate modeling and leads to albedo once summed over wavelengths. For a perfectly Lambertian surface, the hemispherical reflectance is simply equal to the reflectance factor independently of viewing geometry. Lambert's model is frequently implicitly assumed while calculating albedo. Here we are measuring surface reflectance with a near-nadir pointing and various solar incidence angles. We will use an average Mars surface phase function recently derived to account for the non-Lambertian behavior of Mars surface observed at km-sized spatial resolution (Vincendon, 2013): the hemispherical reflectance is typically 10% higher than the nadir reflectance, as Mars surface preferentially scatters light at high emergence angles. This effect is at first order weakly dependent on wavelength at global scale. We do not account for spatial variations of the surface phase function due to the complexity of Mars surface phase function derivation (see e.g. (Fernando, et al., 2013)).

For non Lambertian surfaces, hemispherical reflectance is a function of illumination conditions, i.e. solar zenith angle and aerosols optical depth. As a consequence, there is not "one" albedo for a given place, and albedo maps must be referenced to a given illumination condition. We have selected a solar zenith angle *i* of 45° and a visible aerosols optical depth of 0.7 for two reasons: first, this represents the typical average condition on Mars and second, at *i* = 45° the hemispherical reflectance is nearly independent of the aerosols optical depth (see figure 9 of (Vincendon, 2013)). Albedo maps at these reference values can be simply converted to other illuminations conditions using relations provided in figure 9 of (Vincendon, 2013). The detailed conversion coefficients between nadir reflectance and hemispherical reflectance are provided in figure 9 of (Vincendon, 2013). We use the same optical depth for the conversion as that used for the aerosols correction (see previous section 2.3). To first order, this correction corresponds to a nearly constant shift of about

## 3 Results: validation

### 3.1 Map coverage

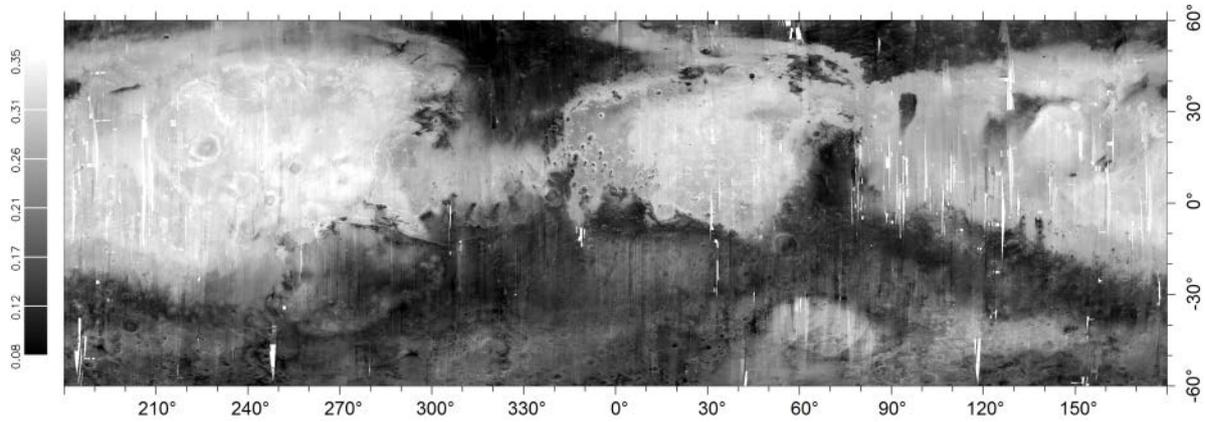

Figure 3: OMEGA albedo map of non-polar latitudes: maximum albedo coverage (98,5%) derived by averaging data from $L_S$ 330° MY 26 to $L_S$ 135° MY 30. The map obtained with quality level #1 is shown in front (96,5% coverage), with gaps filled with the quality level #3 map (see Table 1 for quality level definition). Mismatches between observations are apparent due to surface albedo changes over the period and uncertainty in the aerosols correction.

OMEGA observations used in this study have been obtained from $L_S$ 330° / MY26 (January 2004) to $L_S$ 135° / MY30 (August 2010). Overall, 96.5% of the surface has been observed at least once with high quality observations, and 98.5% if we use also lower quality observations (we consider here non polar latitudes only, from 60°S to 60°N, i.e. more than 85% of the total surface) (Figure 3). During the period of low dust optical depth ("clear season"), the aerosols correction is expected to be more reliable (lower relative uncertainty and lower local dust storm activity). The transition between the clear and storm season typically occurs about $L_S$ 130° and $L_S$ 0° (Figure 1, see also (Wang & Richardson, 2015)). A global dust storm (GDS) started at $L_S$ 265° in MY28 (July 2007) with a maximum visible optical depth of 4 reached at $L_S$ 285°. During decay the optical depth was back to 1.5 at $L_S$ 315° and 0.85 at $L_S$ 340°. Atmospheric dust loading during the first part of the storm season ($L_S$ 130° – 265°) is lower in MY28 compared to MY27 and MY29. We have divided OMEGA observations into 6 periods based on these considerations: we typically separate each year in a clear and dusty season at $L_S$ 130°, except for MY28 (Table 2 and Figure 4). From 25% to 60% of the surface is covered with high quality observations for each period (Table 2).

Table 2: OMEGA observations obtained from January 2004 to August 2010 are divided in 6 periods to separate data gathered when optical depth and storm activity are low, referred as "clear season", from data acquired during the storm (or "dusty") season. MY28 is specific with a lower storm activity between $L_S$ 130° and $L_S$ 265° (compared to MY27 and MY29) and a global dust storm after $L_S$ 265°. The % of the surface covered by OMEGA for the 6 maps are indicated for high and low quality level data as defined in Table 1.

| Period | MY27 clear | MY27 dusty | MY28 | MY29 clear | MY29 dusty | MY30 clear |
|---|---|---|---|---|---|---|
| Start | MY26 $L_S$ 330° | MY27 $L_S$ 130° | MY28 $L_S$ 0° | MY28 $L_S$ 315° | MY29 $L_S$ 130° | MY30 $L_S$ 0° |
| End | MY27 $L_S$ 130° | MY27 $L_S$ 360° | MY28 $L_S$ 265° | MY29 $L_S$ 130° | MY29 $L_S$ 360° | MY30 $L_S$ 135° |
| % cover quality #1 | 39% | 59% | 32% | 28% | 49% | 27% |
| % cover quality #3 | 44% | 72% | 42% | 37% | 51% | 38% |

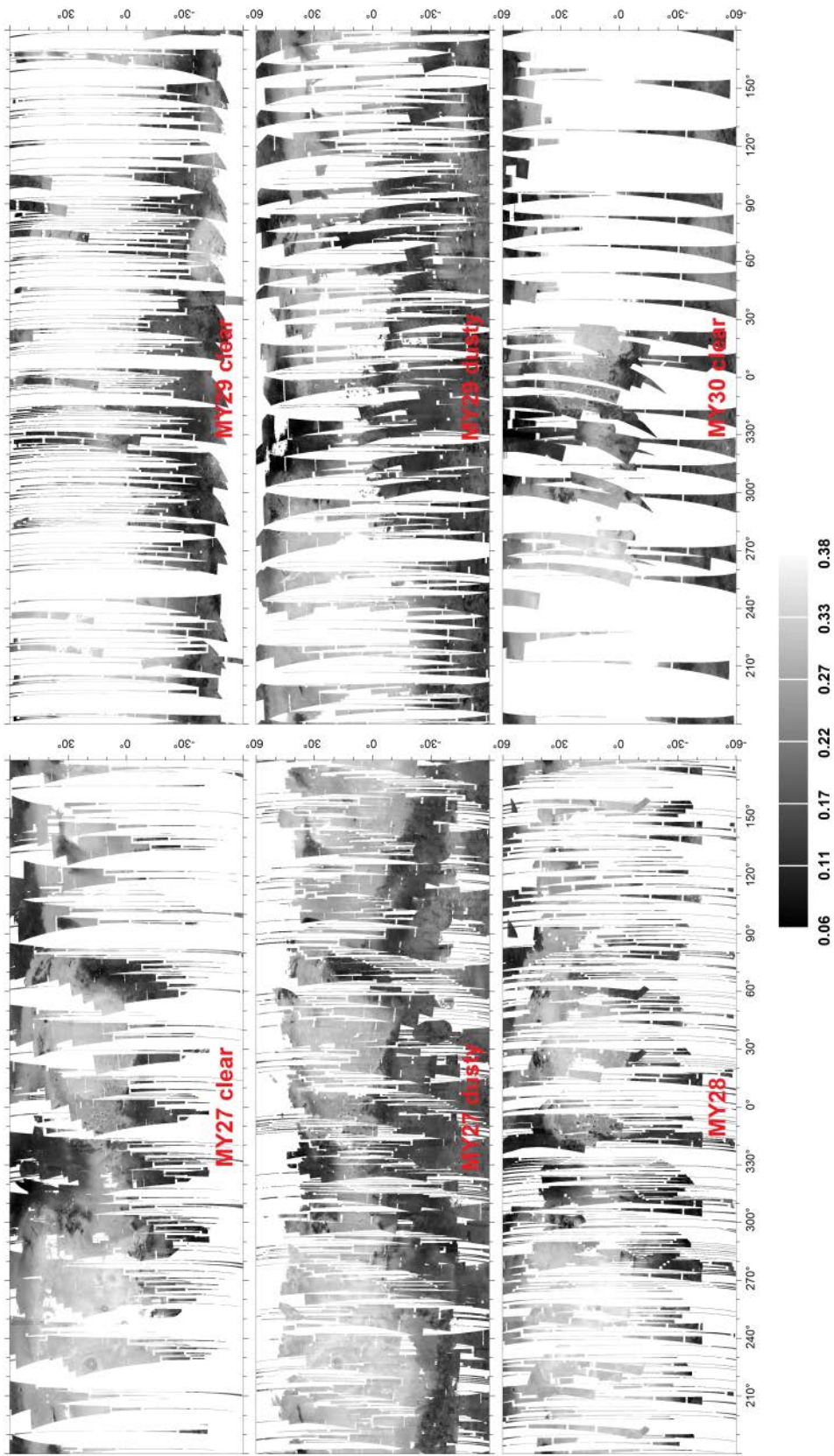

Figure 4: global albedo maps for the 6 periods defined in Table 2 (quality level 1: high quality data only).

## 3.2 Impact of aerosols

As discussed previously (section 2.3), change in the contribution of scattered light by aerosols associated with variations of viewing and atmospheric conditions is a major contributor to apparent albedo change. Uncertainties in the removal of the aerosols contribution can be significant due to uncertainties and variability in the single scattering parameters and optical depth. We link a relative error based on optical depth and particle size uncertainties to each albedo (see 2.3). However, systematic bias may be present in the optical depth absolute level or in the behavior of aerosols with photometric angles (see 2.3). Moreover, local optical depth at time and place of observation are estimated from interpolation/extrapolation schemes which may also lead to non-appropriate values for certain observations.

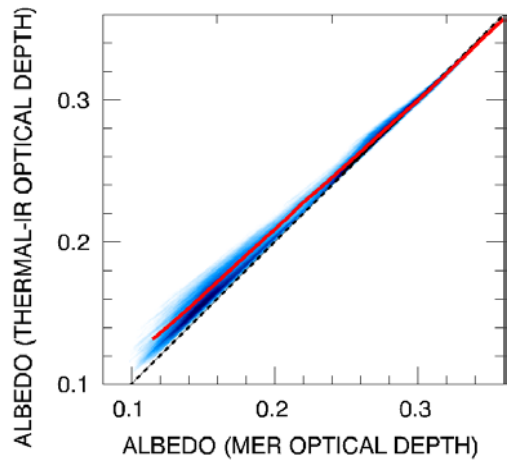

Figure 5: Correlation diagram of surface albedo corrected with the thermal-IR derived optical depths (Montabone, et al., 2015) versus surface albedo corrected with the MER (Lemmon, et al., 2015) derived optical depths (see section 2.3). Data density ranges from dark blue (high) to white (typically 2σ from maximum). The mean value is shown as a red line, and y=x is indicated with a dotted line.

We compare in Figure 5 the albedo obtained with our two optical depth assumptions (see 2.3): the one based on visible surface measurements from the MER (Lemmon, et al., 2015) (hereafter "MER optical depths"), and the one based on thermal-IR orbital mapping (Montabone, et al., 2015) (hereafter "thermal-IR derived optical depths"). The global comparison of all data reveals a systematic difference between both assumptions for low albedo surfaces. Thermal-IR derived optical depths are lower and lead to 20% higher low albedos (Figure 5). As discussed previously (2.3), there are some significant uncertainties in the conversion of thermal-IR optical depths derived from orbit to visible optical depths. In particular, MY27 to MY30 thermal-IR derived optical depths have been obtained from Mars Climate Sounder limb observations which are biased toward lower values according to (Montabone, et al., 2015). On the other hand, MER directly measured the optical depth at visible wavelengths. It has been suggested that Opportunity optical depths may include a slight contribution from water ice aerosols about northern summer solstice (Lemmon, et al., 2015). This putative effect should be of low impact here as we average Spirit and Opportunity measurements and account for a ±20% uncertainty (Figure 1). Moreover, it cannot be systematic over all the year contrary to the observed bias. Additionally, such a contribution from thin water ice aerosols will also contaminate our albedo data (see discussion in section 2.3 about the water ice threshold), and increasing the dust aerosols optical depth in the model would act as a first order removal of thin water ice scattering compared to no correction. As a consequence, the average absolute level of optical depth, and hence the albedo, is expected to be more reliable with MER derived optical depth.

The thermal-IR database is expected to provide a more comprehensive view or the latitudinal and longitudinal variability of optical depth. However, we do not find evidence that using thermal-IR derived optical depth does provide a general better match between observations than using MER optical depths. While the use of thermal-IR optical depths does improve correction for some observations, it worsens correction for others. For example, we compare in Figure 6 mosaics obtained with both optical depths for a situation of expected better performances of thermal-IR derived optical depths: we analyze an area located far from MER landing sites at 137°E, 31°N and compare data acquired during the storm season where regional dust events are frequent. This area includes strong albedo contrast with dark albedos very sensitive to aerosols scattering. We can see that mismatch between observations are only partly accounted for by the MER derived optical depths, as expected. However, we can see that the correction is overall not improved with thermal-IR derived optical depths: one observation better overlaps, while artifacts appear for two observations. This may result from insufficient spatial and/or temporal sampling in the (Montabone, et al., 2015) database. A new version of the database, derived after we perform this study, now includes a reliability factor associated with interpolation, which may help removing these artifacts.

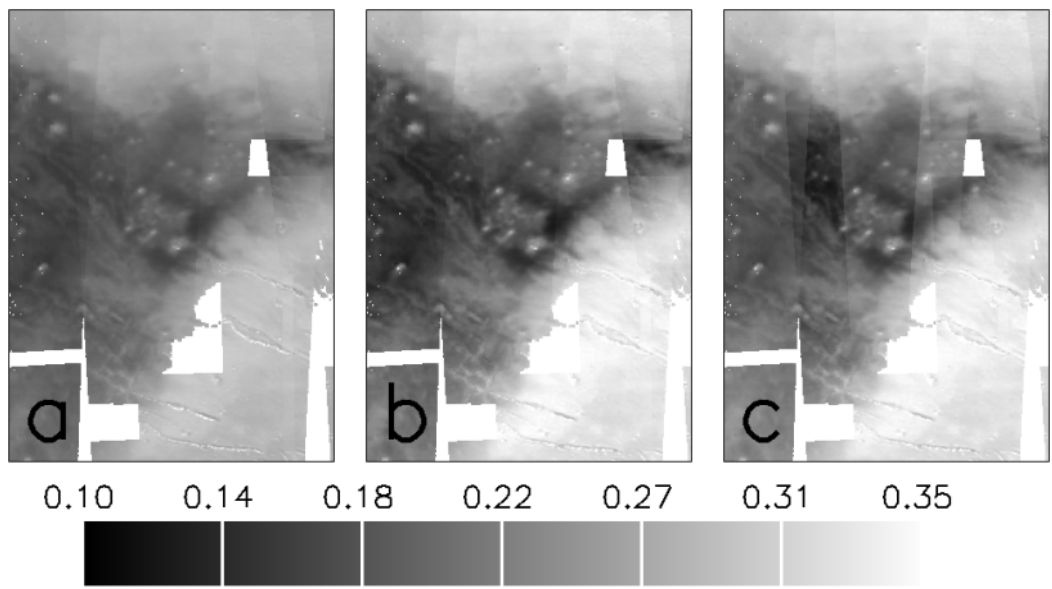

Figure 6: Comparison between raw albedo map (a), albedo map corrected with MER optical depths (b), and albedo map corrected with thermal-IR derived optical depths (c) in an area located far from MER landing sites (137.2°E, 31°N). Image is 9.4° x 13.1° wide. The mosaics is built with high quality data (level 1, see Table 1) acquired during MY27 dust season (see Table 2) when localized events are expected.

From this analysis, we reach two conclusions. First, the MER derived optical depth will be used as our default scenario: the absolute level of optical depths is more reliable and the use of thermal-derived maps does not generally increase the quality of match between various observations. Second, we will carefully consider potential aerosols biases during our analysis when a surface variability is suspected: coherence between successive observations, correlations with photometric conditions, optical depth assumption. For example, we observe a low albedo feature 100 km wide at 279°E and 48°N during the storm season ($L_s$ 185°), not apparent in previous ($L_s$ 124°) and following ($L_s$ 46°) overlapping observations. A careful look at the images reveals a dark feature that changed between two observations taken 3° of $L_S$ apart (observations # 6428_5 at $L_S$ 185° and # 6446_5 at $L_S$ 188°): this feature corresponds to the shadow of a local dust storm.

### 3.3 Comparison with Thermal Emission Spectrometer (TES)

The reference albedo maps currently widely used for Mars studies are those obtained by the Thermal Emission Spectrometer (TES) instrument (Christensen, et al., 2001). TES provided bolometric reflectance measurements with a point spectrometer over 0.3 – 2.9 µm. TES was onboard Mars Global Surveyor which orbit was heliosynchronous (14.00 local time) and circular (constant 3 km pixel footprint). TES measurements of surface bolometric solar reflectance were not corrected for atmospheric or surface effects but clear-condition observations were selected to build the maps. Moreover, the small solar zenith angles of TES early afternoon observations were optimum to minimize aerosols effects, as well as the ability of TES to detect water ice clouds (Smith, 2004). Several Mars Years have been observed by TES; albedo maps with a partial (but regularly distributed) coverage have been derived from MY24 to the beginning of MY27 (Putzig & Mellon, 2007). OMEGA and TES have been observing the surface simultaneously during the first half of MY27, a period corresponding to clear atmospheric conditions. We compare the albedos values of all areas commonly observed by TES and OMEGA in Figure 7, and illustrate this common coverage with a zoom on Syrtis Major in Figure 8. We use an OMEGA map built with observations gathered between $L_S$ 330° (late MY26) and $L_S$ 130° (MY27, end of the clear season), i.e. data coming from almost the whole year 2004. We exclude from this map high resolution OMEGA observations (16 pixels wide tracks) to compare similar surface resolution observations. This map is comparable to the "MY27" TES map provided by N. Putzig (personal communication), which is essentially built with data gathered by TES in 2004, with some additional data corresponding to MY27 storm season (year 2005) used – with a security opacity threshold – to complete some gaps. Overall, we found a very good agreement between TES and OMEGA "raw" albedo values, i.e. between similar albedos that have not been corrected for aerosols or surface BRDF: we just observe a constant shift of +3% in OMEGA albedo compared to TES. We analyze and quantify hereafter potential expected differences between both instruments/methods that could explain this shift:

- Above all else, the difference may simply result from imperfections in instruments calibration, as 3% is below expected relative instrumental precision of TES and OMEGA: absolute accuracy is observed to be better than 10% for OMEGA (see 2.1) and estimated to be about 1-2% for TES visible/near-IR bolometer (Christensen, et al., 2001).
- The broadband channel of TES includes two significant $CO_2$ gas absorptions, between 1.9 µm and 2.1 µm and between 2.65 µm and 2.9 µm, while we correct or exclude these bands with OMEGA spectral measurements. This artificially reduces TES albedo by about 1.5% (the main contribution is due to the 2 µm band).
- Our wavelength range starts at 0.25 µm instead of 0.3 µm for TES. TES albedo thus includes a smaller contribution of dark UV reflectance, which this time artificially increases TES albedo by 1%.
- Potential reflectance differences due to surface phase effect are typically a few % for nadir viewing observers (Vincendon, 2013). However, surface phase effects are not expected to result in a systematic shift here, as OMEGA varying local time results in either smaller or higher solar zenith angle compared to TES constant 14:00 local time (both maps have been acquired during the same season, see above).

- As explained previously (section 2.2)., assumptions and simplifications during the interpolation procedure can result in 1% albedo variations for OMEGA.

To conclude, differences in observation methods (wavelength range, atmospheric gas absorption, solar zenith angle) do not seem to easily explain alone the 3% systematic shift, which is probably at least partly due to a difference in absolute calibration fully compatible with expected instruments performances.

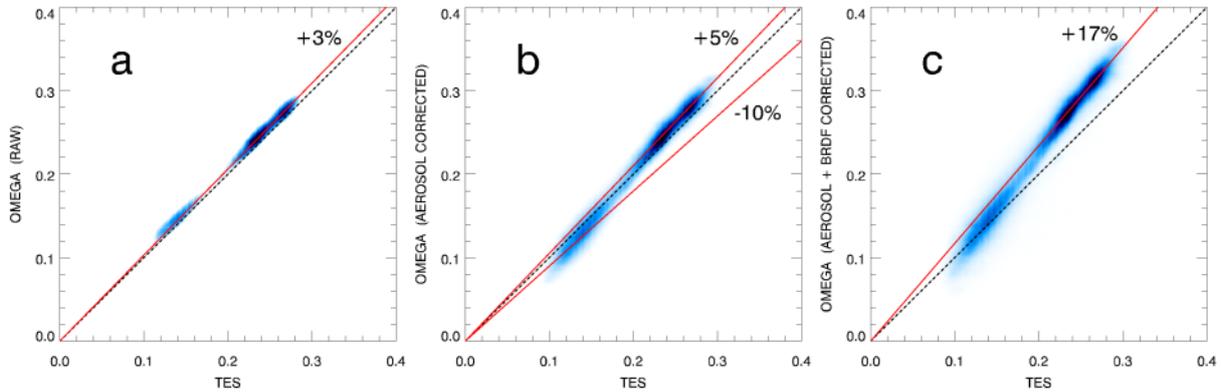

Figure 7: Comparison between OMEGA and TES albedo. Data density ranges from dark blue (high) to white (typically 2σ from maximum). Trend lines are indicated in red. TES albedos correspond to bolometric reflectance measurements not corrected for aerosols or surface BRDF. Corresponding OMEGA raw albedos are compared on panel (a): OMEGA raw albedos are on average 3% brighter than TES albedo. Aerosols corrected albedos are shown on panel b: the aerosols correction significantly reduces dark albedo (-13%, i.e. -10% compared to TES) and increase bright albedos (+2%, i.e. +5% compared to TES). Finally, hemispherical albedos also corrected for surface BRDF are compared to TES on panel (c). The surface BRDF correction essentially results in a global shift by 10% of albedo. As a result of all corrections, OMEGA dark albedos are very similar to TES, while bright albedos are +17% brighter.

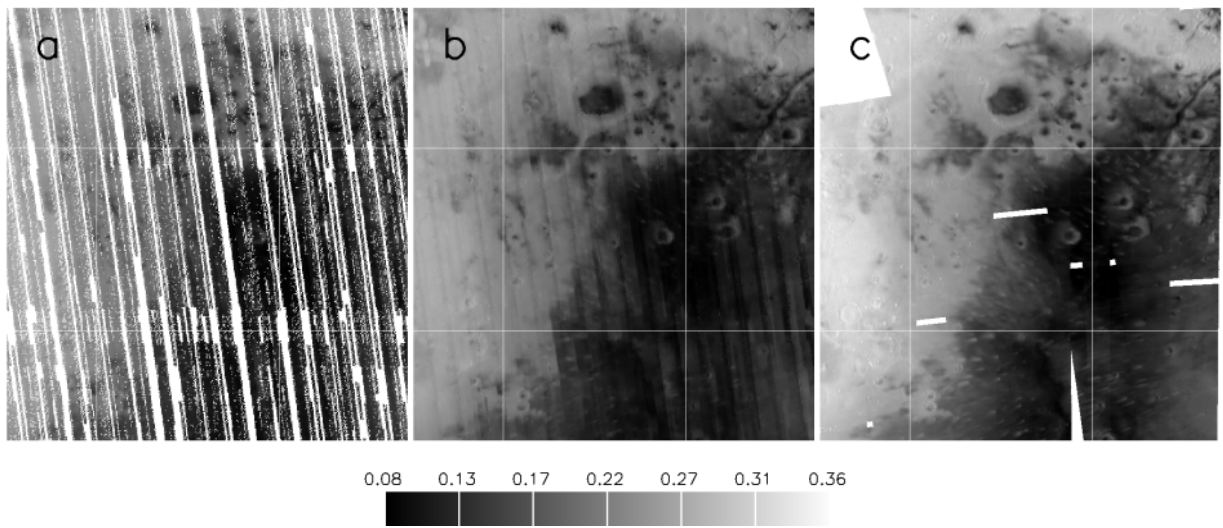

Figure 8: comparison of TES and OMEGA albedo maps of Syrtis Major obtained during the same period (MY27 clear season). (a) TES map, (b) TES interpolated map, (c) OMEGA map. TES MY27 maps provided by N. Putzig (see also (Putzig & Mellon, 2007)).

We also compare in Figure 7 TES albedo with OMEGA processed albedo (i.e., with aerosols and surface photometry correction). Aerosols correction increases the contrast of OMEGA albedo compared to TES: dark areas are now 10% darker while bright areas are 5% brighter. Accounting for surface photometry then roughly shifts all values up, so that dark albedos are similar to TES values while bright albedos are +17% brighter. Most differences between TES and OMEGA albedo thus result from atmospheric and surface photometry corrections applied rather than intrinsic differences between both instruments. We finally compare in Figure 8 OMEGA and TES maps of Syrtis major obtained during the same MY27 clear season. Over a given season, TES (3 km footprint) gathered regularly spaced data of the surface with gaps between observation swaths. Complete maps can be obtained with spatial interpolation that is typically equivalent to a spatial sampling decrease by a factor of two. OMEGA provides continuous coverage of observed places with spatial sampling about 2 km. Gaps between observations are not regularly spaced and can be significantly larger than the spatial resolution.

# 4   Results: surface change

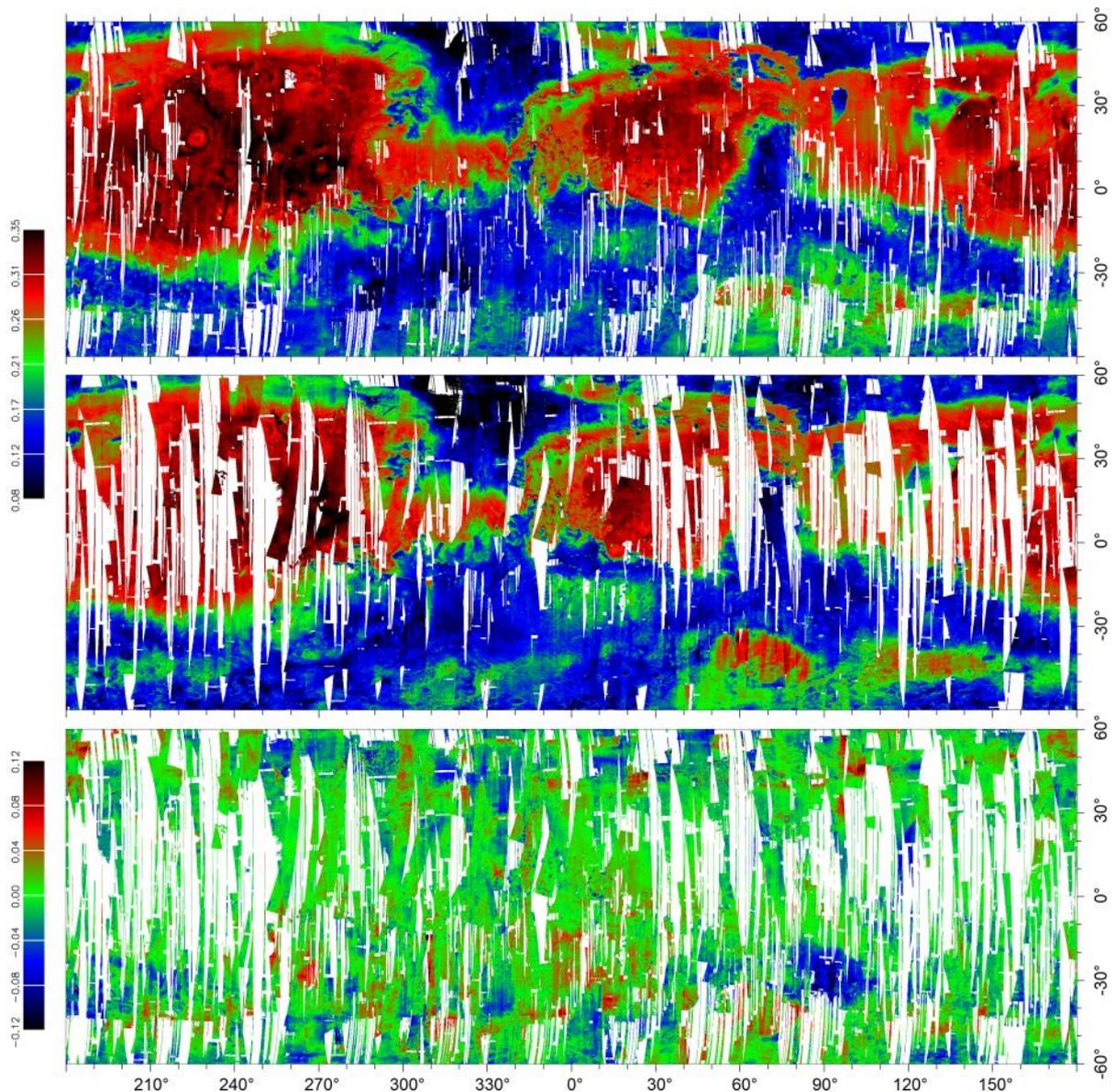

**Figure 9: (top) albedo map prior to the MY28 GDS (orbits ≤ 4463: data from MY26 LS 330° to MY 28 L$_S$ 265°: 86% coverage). (middle) albedo map after the MY28 GDS (orbits ≥ 4758: data from MY 28 L$_S$ 315° to MY 30 L$_S$ 135°: 74% coverage). (bottom) difference albedo map (middle – top: 63% coverage). Quality level # 1 (Table 1) is used. Bright areas or brightening ≥ 0.04 are in red, dark areas or darkening ≤ -0.04 are in blue. Intermediate albedo and stable areas are in green.**

Differences are observed between seasonal albedo maps of Figure 4. Major albedo changes are linked with the MY28 GDS, as observed during the previous MY25 GDS (Szwast, et al., 2006; Cantor, 2007). We compare mosaics of pre and post GDS observations on Figure 9, which reveal albedo change by more than ± 0.04 over about 9% of the surface (63% of the surface have been observed at least once before and after the GDS). A diversity of combination of seasons is present in this comparison: post-GDS observations can be obtained either directly after the GDS or more than one year later; both clear season and storm season observations are included in the maps; etc. While the direct interpretation of such a global map is tricky, it highlights some major changes (or lack of changes). We can see that most changes occurred in places with dark or intermediate albedos, and that brightening and darkening occurred nearly equally. A similar situation was observed after the 2001 GDS (Szwast, et al., 2006), but changes are not identical. We zoom on changes at several typical locations in the next paragraphs (Figure 10).

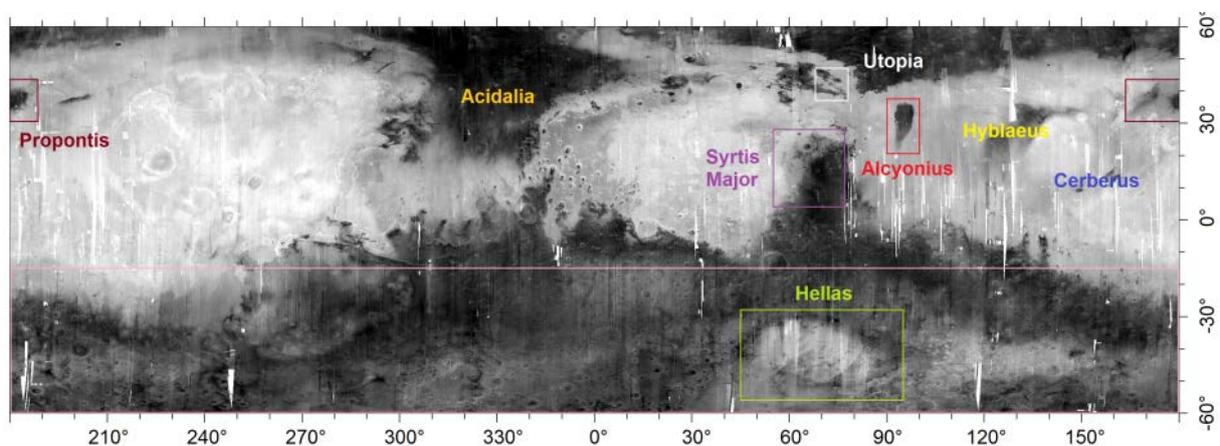

Figure 10: Albedo map with name and location of regions discussed in details in section 4. Boxes correspond to Figures #11 (red), #13 (white), #14 (brown), #17 (purple), #18 (pink) and #19 (green).

## 4.1 Dark northern plains: Utopia and Acidalia

Observations from last century reveal changes in and near the dark northern plains Utopia, north Arcadia, and Acidalia. The dark areas located near the bright/dark margin south of Utopia Planitia (around 40°N and 80°E), such as Alcyonius or Nilosyrtis, can be alternatively cleaned (1954, 1999) or partly buried (1972, 1978, 2001) (Geissler, 2005; Szwast, et al., 2006). These dark areas, as well as the southern part of Acidalia (35°W, 25°N), enlarged between Viking and HST or Mars Global Surveyor observations obtained about 20 years later (Bell, et al., 1999; Geissler, 2005). These modifications were interpreted as evidence for a progressive cleaning caused by wind or dust devil erosion over several decades (Geissler, 2005). However, such a steady process was not apparent in successive 1999-2004 MGS data which showed more irregular modifications caused by rapid local deposition or removal of dust linked with regional or global dust storms (Szwast, et al., 2006).

We observe significant changes (both brightening and darkening) in the dark northern plains after the GDS, notably near the southern dark / bright margins (Figure 9). Overall, changes observed near the boundary between dark northern plains and bright southern terrains are more extensive after MY28 GDS compared to MY25 GDS (Szwast, et al., 2006). Most changes occur near the margin between bright and dark terrains, thus resulting in apparent northward movement of bright terrains or southward movements of dark terrains. Brightening is observed at 190°E, 50°N east of Utopia, while darkening occurred after MY25 GDS (and then brightening the following year) (Szwast, et al., 2006). Brightening is observed near 105°E (south Utopia), while darkening was observed there between Viking and MGS (Geissler, 2005). Darkening is observed eastward at 125°E. Thus, we do not observe evidence for an unvarying and widespread steady trend in frontier movements, contrary to previous expectations (Bell, et al., 1999; Geissler, 2005). Our observations are however not inconsistent with possible local and episodic boundaries advances (Geissler, et al., 2013). Changes also occur inside dark areas far from albedo frontiers, notably darkening in Acidalia/Chryse (Figure 9), where dust lifting is known to be extensive (Cantor, et al., 2001; Basu, et al., 2006; Wang & Richardson, 2015). We detail observed changes in two typical areas in the next paragraphs.

### *4.1.1 Alcyonius*

The Alcyonius dark feature have been shown to significantly enlarge southward between 1978 and 1980, and then between 1980 and 1999 (Geissler, 2005). Detailed observations collected over 1999-2004 (Szwast, et al., 2006) show no change during MY24 and MY25 prior to the MY25 GDS. Bright material was deposited during the GDS in the southern portion of the feature. This dust persisted up to $L_s$ 210° the following year (MY26). A regional dust storm then removed dust in the southern part of Alcyonius. As a result, "the southern boundary of the feature had become more elongated and sharpened" in late MY26 / early MY27 compared to MY24 (Szwast, et al., 2006).

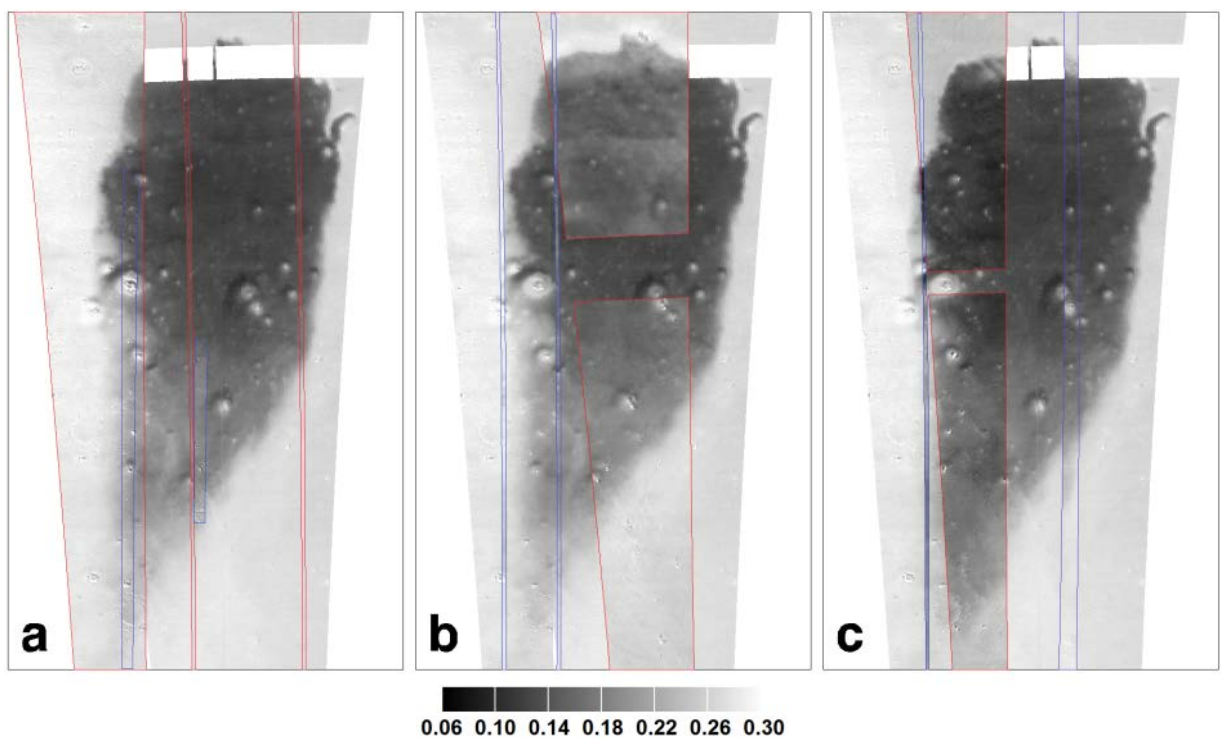

Figure 11: OMEGA albedo observations of Alcyonius. Images are 10 x 17° wide, centered on 29°N and 95°E; north is on top. (a) MY27-28 map. Background: MY27 clear season ($L_s$ 21° and 127°). Blue outlined tracks: MY27 storm season ($L_s$ 272°). Red outlined tracks: MY28 before the GDS ($L_s$ 72° and 245°). (b) Post MY28 GDS observations (outlined) shown on top of a MY27-28 background. Blue outlined: MY29 $L_s$ 38°. Red outlined: MY29 $L_s$ 187°. (c) MY30 clear season (outlined) shown on top a MY27-28 background. Blue outlined: MY29 $L_s$ 336° and MY30 LS 10°. Red outlined: MY30 $L_s$ 126°.

We observe that the area remains essentially stable over MY27 and MY28 (Figure 11). Observed surface albedo variations (typically ± 0.01) are within aerosols uncertainty error bars (Figure 12). The MY28 GDS modified the Alcyonius area in several ways depending on location. Dust was deposited over all the northern (darker) half of Alcyonius with albedo increase typically from 0.13 to 0.18. In the southern part of the feature, the already higher albedos (0.20) were not modified by the GDS. Significant darkening (e.g., from 0.24 to 0.16) is observed in the bright west frontier (Figure 11) right after the GDS. Thus, some dust was removed from the south west bright part of Alcyonius and some dust was deposited in the north dark part during the same GDS. Deposited dust persisted at the surface up to at least $L_s$ 202° / MY29. Then, these areas get cleaned between $L_s$ 202° and the next available observation at the end of the storm season ($L_s$ 336°). The albedo remains low and unchanged during MY30 clear season up to $L_s$ 126° at a slightly lower value than prior to the GDS although differences might be due to aerosols correction uncertainty (Figure 12). Dust cleaning also occurs in areas not affected by the GDS in the south of the feature: the Alcyonius feature appears more elongated southward in MY30 than it was in MY27. We thus observe a continuation of the southward enlargement of the dark feature observe between 1978 and 2004 by (Geissler, 2005; Szwast, et al., 2006). This long-term growth trend results from irregular cleaning events occurring during storm seasons after $L_s$ 200°. As change appears from one observation to the next, the most likely mechanism is punctual massive dust removal due to storms, as favored by (Szwast, et al., 2006). However, the time gap between two OMEGA observations can be large and we cannot rule out the action of a gradual process such as dust devil cleaning (Geissler, 2005; Geissler, et al., 2013). The darkening trend is not a regular process as it breaks off due to temporary brightening events. Brightening also appears to occur as isolated events during storm seasons. Over 1999-2010 (7 mars years), we observe that various events can occur in a given storm season (darkening or brightening, in the north or in the south), without apparent regularity. A similarity is apparent between the MY24-MY27 and MY27-MY30 years: major changes occurred during the GDS or during the storm season following the GDS.

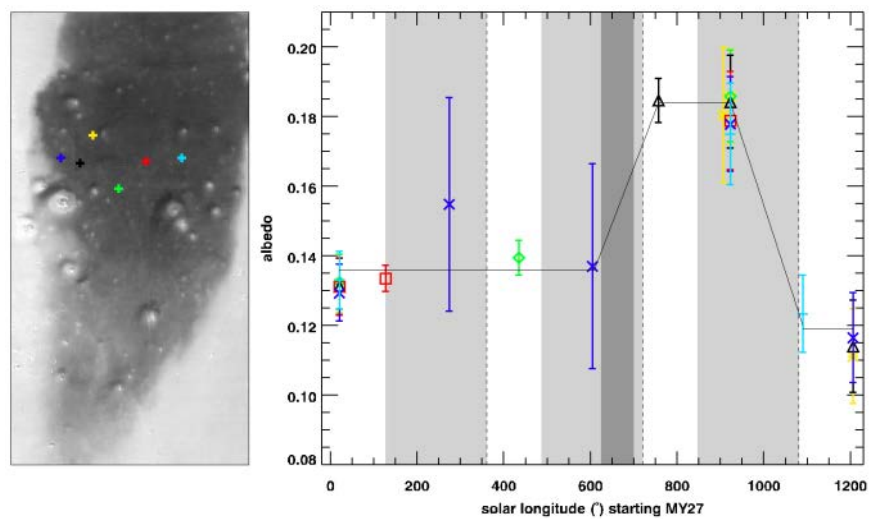

Figure 12: Time evolution of the albedo of north Alcyonius over MY27-MY30. We compare data collected over 6 similar positions to increase time sampling. Positions are mapped with the same colors. Relative error bars correspond to the aerosols correction as discussed in section 2.3. Storm season ($L_s$ 130° - 360°) are indicated in light grey. The MY28 GDS is indicated in dark grey. The trend consistent with error bars is indicated as a thin line.

### *4.1.2 South west Utopia*

Our second region of interest is located 1000 km north west from Alcyonius (Figure 13): it is composed of irregular dark areas located south of the main Utopia bright/dark boundary. A previous MGS (1999-2004) analysis showed that these areas were stable during MY24 and MY25, largely covered by an optically thick cover of dust at $L_s$ 309° after the decay of MY25 GDS, and then essentially restored by $L_s$ 73° / MY26 probably due to storms activity between $L_s$ 309° and 328° in MY25 (Szwast, et al., 2006). Locally, some dust cover remained without change during MY26 and the beginning of MY27 (Szwast, et al., 2006). We observed that the area is stable during MY27 and MY28 prior to the GDS (Figure 13). No observations were obtained right after GDS decay ($L_S$ 340°). Subsequent MY29 observations at $L_S$ 30°-45° reveal an area still recognizable in shape but largely covered by an optically thin dust layer with albedo increase by about 0.1 for dark places (+50% to 100%) and by 0.05 for bright places. No evolution is seen between $L_S$ 30°-45° and $L_S$ 101° (clear season). While the area was already largely restored during the clear season following MY25 GDS, this is thus not the case after the MY28 GDS. This may be due to the different timing between the two GDS, as the MY28 GDS ended at $L_S$ 340°, after the period of suspected cleaning activity in MY25 ($L_S$ 309° - 328°). The area then appeared partly cleaned at $L_S$ 190 - 200° (storm season of MY29), and even more cleaned at $L_S$ 350° (end of storm season), with about half of the albedo increase that has disappeared. Other bright areas not modified by the MY28 GDS are also cleaned during MY29. The remaining brightening (+ 0.05 for dark places) is observed to persist during the MY30 clear season. An adjacent area covered by dust during the MY25 GDS (Szwast, et al., 2006) (see figure 31 of this reference) has also still not been restored in MY30 in our dataset (Figure 4, Figure 13).

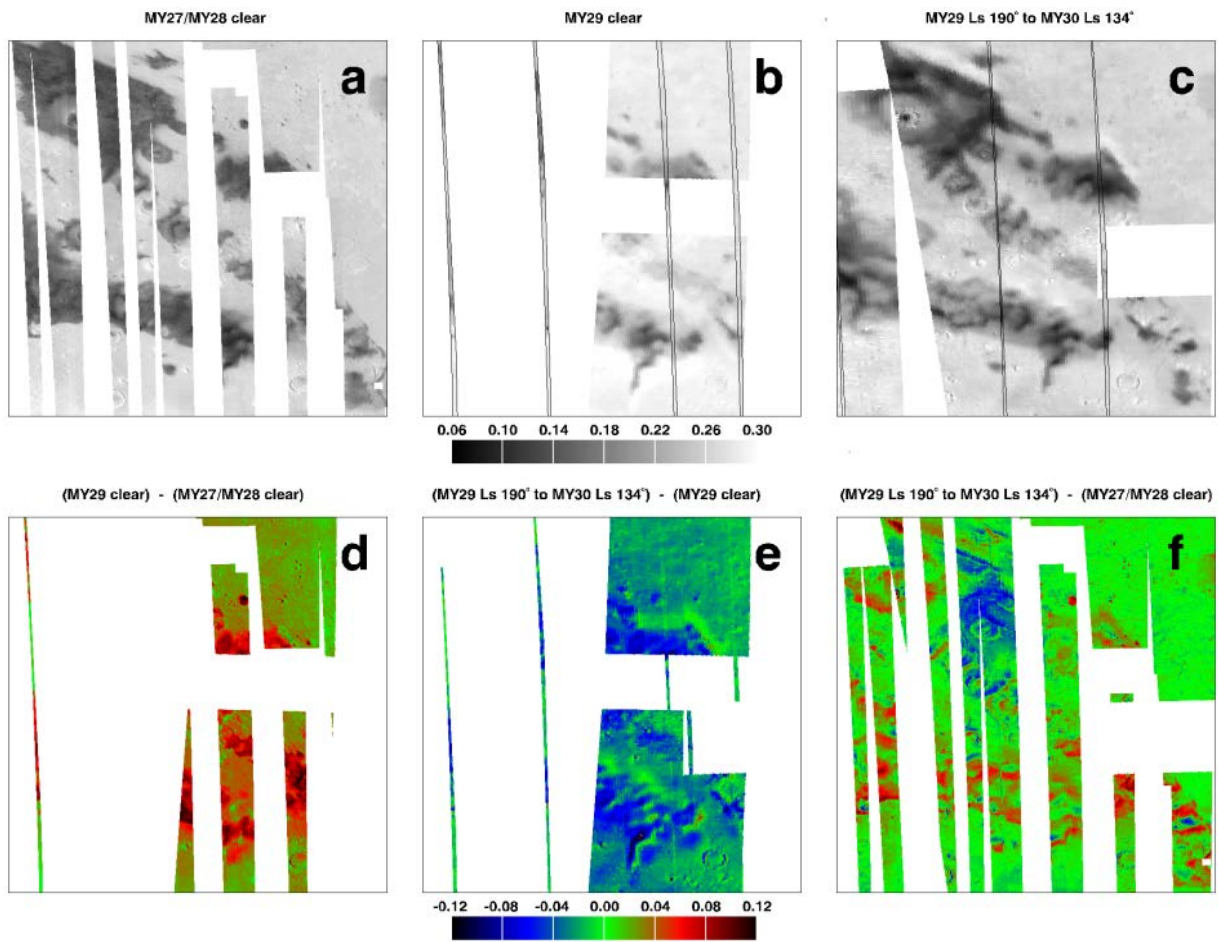

**Figure 13: Deposition of dust at the South west frontier of Utopia after the 2007 global dust storm. A 10x10° wide area centered on 73°E and 42°N is shown (north on top). Top (a, b, c): albedo maps. (a) clear seasons of MY27 and MY28; (b) clear season of MY29 : $L_S$ 101° in background and $L_S$ 30-45° for the four thin outlined tracks; (c) storm season of MY29 and clear season of MY30 (MY29 $L_S$ 190-200° left and middle, MY30 $L_S$ 134° right, MY29 $L_S$ 350° for the thin outlined tracks). Bottom (d, e, f): corresponding albedo difference maps. Albedo increases are in red, stable albedo in green, and albedo decreases in blue: (d) dust deposited during the MY28 GDS; (e) MY29 storm season cleaning; (f) summary 2004 / 2010. While the area is stable during MY27 and MY28 pre-GDS (a), dust deposits caused by the MY28 GDS (d) are observed to**

**persist during all the clear season of MY29 (b). A partial cleaning then occurred during the storm season of MY29 (c, e). On the whole, modifications arisen between 2004 and 2010 resulted in a complex pattern of changes in the area (f).**

To conclude, dust is deposited during GDS and cleaned during usual storm seasons. These changes are not regular: some dust deposited during the 2001 GDS had still not been removed in 2010, while dust can be removed from a place while not being deposited there during the last GDS. There are no apparent trend between 2004 and 2010 in that area, which is well summarized by Figure 13f where we can see that both brightening and darkening occurred when we compared clear season maps of MY30 to MY27. We can also notice that the changes are small-scaled, while the overall shape of the feature is very similar in 2010 compared to its appearance at the beginning of MGS mission in 1999.

## 4.2 Around Elysium Mons

### *4.2.1 Propontis*

The bright Arcadia province includes a dark feature near 180°E and 35°N named Propontis. The region darkened between Viking late 70s observations and telescopic or orbital observations obtained about 20 years later (Bell, et al., 1999; Geissler, 2005). Then, the dark feature was cut in two parts (west and east) as a layer of bright material taken to be dust gets deposited in its center in 2003 (Szwast, et al., 2006). The dust layer was probably deposited during a single event related to a regional dust storm that developed over the area at $L_S$ 213° in MY26 (Szwast, et al., 2006). The dust layer was still observed at $L_S$ 10° the following year (Figure 14).

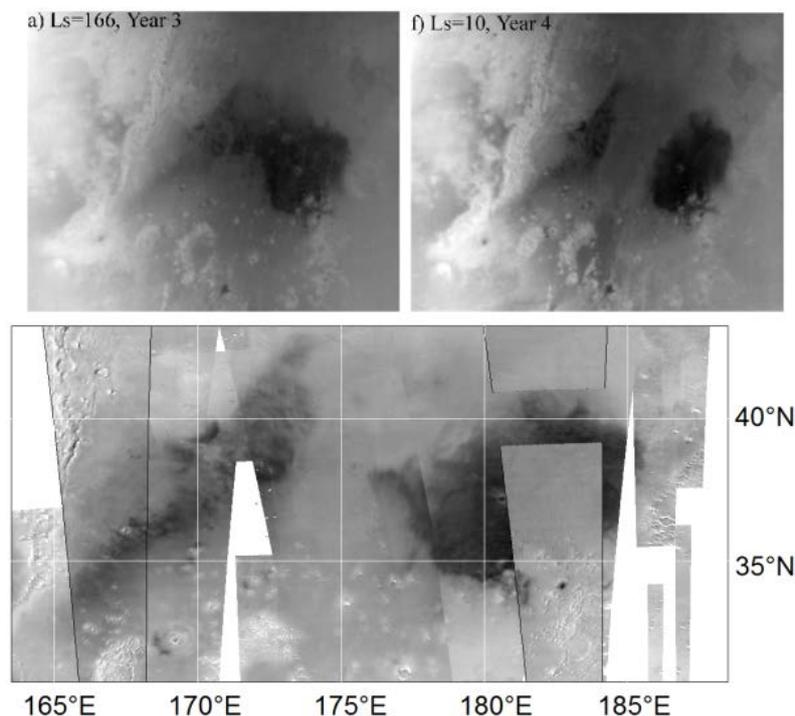

Figure 14. Observations of Propontis. Top: two observations from (Szwast, et al., 2006) that corresponds to $L_S$ 166° of MY26 and $L_S$ 10° of MY27. Bottom: mosaic of OMEGA observations (background: $L_S$ 15-26° and 284-341° of MY27, $L_S$ 260-264° of MY28 and $L_S$ 185° of MY29); black outlined: $L_S$ 131°-135° of MY30).

Minor changes are observed during MY27 and MY28 considering the aerosols uncertainties, with essentially movements of the margin between the new central bright area and the east dark feature (Figure 15): the dust cover extended eastward in the beginning of MY28 compared to earlier

MY27 and later MY28 observations. The first observations obtained at $L_S$ 12° after the MY28 GDS reveal no changes in bright/dark frontiers or dark albedo level compared to observations obtained just before the onset of the GDS. Thus, the GDS did not result in any changes (dust removal or deposits) in that area. As observed previously around volcanoes (Cantor, 2007), winds must be keeping the GDS fallout from being deposited in Propontis. Significant activity is observed during the MY29 storm season, with substantial brightening at $L_S$ 200° and especially at $L_S$ 335° with albedo values of 0.3 – 0.4 above the dark area suggesting strong local atmospheric dust scattering (Figure 15d). Observations obtained during the clear season of MY30 at $L_S$ 5° and $L_S$ 135° then reveal a persistent dust cover that masks the east dark part of the area (see also (Vincendon, et al., 2013)), while no changes are observed in the west dark part (Figure 14). A few dark spots south of the east feature remained. The new albedo (0.22-0.24, versus about 0.10-0.14 previously) is similar to some directly surrounding areas, and lower than the 0.3 value found in the brightest nearby areas and in areas associated with thick dust cover such as Tharsis. The dust deposited in the central part of Propontis during the storm season of MY26 is still observed in our latest MY29 or MY30 observations. Mars Color Imager (MARCI) observations revealed that the MY29 change probably essentially occurred between $L_S$ 322° and 327° when regional dust storms where observed over the area (Lee, et al., 2014), which is fully consistent with the $L_S$ 325° OMEGA observation of Figure 15d. MARCI also showed that the dust cover has been stable at least up to 2014 (MY32), and that the former dark feature shape can still be perceived through the bright material layer (Lee, et al., 2014).

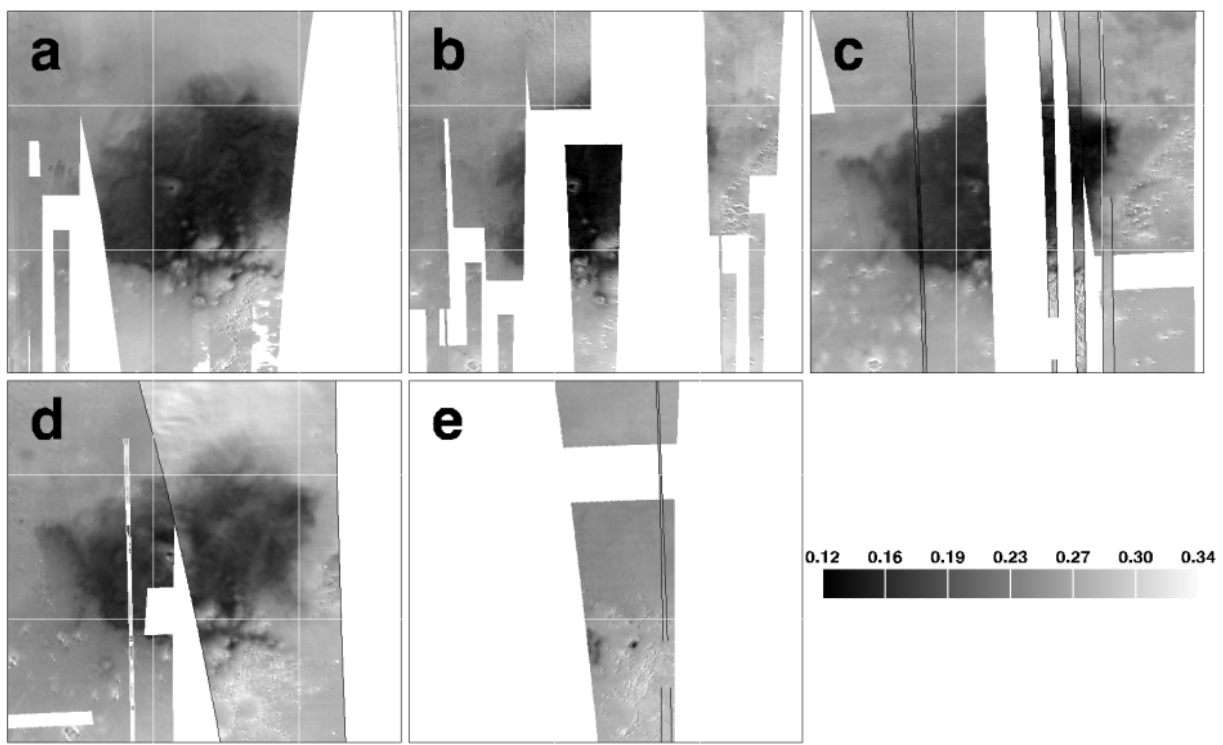

Figure 15: zoom on the east part of Propontis (see Figure 14 for context). (a) Data collected between MY26 $L_S$ 334° and MY27 $L_S$ 182°. (b) Data collected between MY27 $L_S$ 274° and MY28 $L_S$ 84°. (c) 4 outlined observations obtained after the GDS between $L_S$ 12° and $L_S$ 53° in MY29 are placed on top of a background map acquired between $L_S$ 250° and $L_S$ 264° in MY28, prior to the GDS. (d) MY29 storm season observation, with background observations obtained at $L_S$ 184°, black outlined observations obtained at $L_S$ 201° and white outlined observations obtained at $L_S$ 325°. (e) MY30 observations at $L_S$ 5° (thin, outlined, placed on top) and $L_S$ 135° (large).

All these observations point toward the deposition of a bright dust layer not optically thick over the darker unit, which results in an intermediate albedo. Previous estimates of dust thickness associated with such albedo changes are between a few microns to a few tens of μm (Christensen, 1988; Cantor, 2007; Kinch, et al., 2007). OMEGA observations can also be used to constrain the thermal inertia (Audouard, et al., 2014). Before the albedo change, early afternoon observations reveal thermal inertia of about 250-350 J $m^{-2}$ $s^{-1/2}$ $K^{-1}$ in the dark places, while lower values of 120 – 150 J $m^{-2}$ $s^{-1/2}$ $K^{-1}$ consistent with thermally thick (more than 4 cm) dust deposit (Audouard, et al., 2014)) are observed in brighter nearby areas. After albedo change, we do not observe modification of the thermal inertia which is still 250-350 J $m^{-2}$ $s^{-1/2}$ $K^{-1}$ in the newly brightened area. This implies a deposited dust layer thinner than 100 μm (Cantor, 2007; Audouard, et al., 2014). The lack of thermal inertia modification linked with surface albedo change was already observed (Lee, 1986). Our observations are consistent with previous interpretation of these intermediate albedo surfaces with high thermal inertia as being composed of a dark unit covered by a thermally thin dust cover (Geissler, 2005), while not consistent with an optically thick layer of material of intermediate albedo by itself, such as duricrust (Putzig, et al., 2005).

We study in Figure 16 how this not optically thick layer of dust modifies apparent surface mineralogy as detected from orbit with near-IR spectroscopy. In previous OMEGA mineralogical mapping (Ody, et al., 2012), the dark east area of Propontis has distinct spectral properties indicative of the presence of pyroxene with lower amount of ferric oxides compared to surrounding brighter areas which have spectral properties typical of dust. The brightening at the end of MY29 results in a drastic change of the dominant mineral detected: signatures of basaltic rocks (pyroxene) detected prior to the event disappeared, while ferric oxide signatures associated with Martian dust now dominate. This shows that a thin, micrometers to tens of micrometer thick layer of dust is enough to mask mineral signatures of the underlying material. Surface minerals maps derived in the near-IR are thus representative of the very surficial upper layer (microns to tens of microns thick).

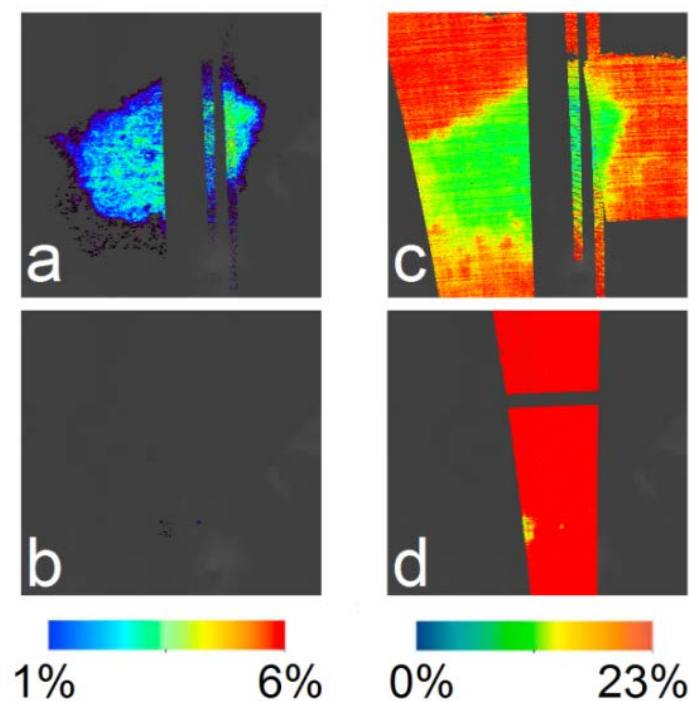

Figure 16: Near-IR mineralogy of the Propontis east dark feature (see Figure 15 for context) before (a, c) and after (b, d) the coverage of the area by dust in late MY29. Spectral parameters described in (Ody, et al., 2012) are shown: (a, b) Pyroxene band depth, with a detection threshold at 1% (no pyroxene detected after the coverage by dust); (c, d) Ferric oxide band depth.

*4.2.2 Hyblaeus*

The dark feature located west of Elysium (Hyblaeus, near 120°E, 25°N) extended southward and westward after MY 28 GDS (Figure 9). While changes were limited to margins over 1999-2004 (Szwast, et al., 2006), with no major impact of the MY25 GDS, significant changes were reported between Mariner and Viking and then between Viking and 1995-2000 observations (Bell, et al., 1999; Geissler, 2005): the south west area of the dark feature was uncovered by dust only during Viking observations, and cleaning occurred in the north east area between Viking and 1995. After the MY28 GDS the north east area is still as dark as during the MGS mission, while the south west area has been cleaned again and is now closer to its Viking appearance. This confirms the supposition that changes in Hyblaeus are not due to seasonally varying winds but are associated with dust storms (Szwast, et al., 2006). Overall, while changes are not similar from one GDS to the next (very little change in 2001, large scale cleaning in 2007), the area appears to alternate between reproducible shapes over decades.

*4.2.3 Cerberus*

The Cerberus 1500 x 500 km dark area was observed to be fairly constant in appearance for a century until it disappeared, possibly during a single event in 1994, or progressively between Viking and 1994 (James, et al., 1996; Bell, et al., 1999; Erard, 2000). No change was then observed between 1999 and 2004 in that area (Szwast, et al., 2006). No major modifications are also found in that area over 2004-2010 (Figure 4, Figure 9): the area is still covered by dust except in the same few dark spots as previously. The fact that the area was first observed to be constantly dark prior to 1980 and then constantly bright after 1994, without robust evidence for a progressive change in between, argues in favor of a brightening resulting from one isolated event rather than from a progressive process.

## 4.3 Syrtis Major

Both seasonal and secular changes of albedo level or boundaries have been reported at Syrtis Major over the last century (Sagan, et al., 1972; Capen, 1976; Christensen, 1988; Geissler, 2005; Szwast, et al., 2006; Cantor, 2007). Changes have been alternately observed essentially in the eastern (Sagan, et al., 1972) or western (Geissler, 2005) margins. During years without global dust storms, the area is observed to be relatively stable, with slight changes at boundaries (Bell, et al., 1999) and in average albedo level (Cantor, 2007). After Viking second GDS that ended late in the storm season at $L_S$ 340°, Syrtis remained slightly brighter (albedo about 0.16 versus 0.13) and got progressively cleaned up to $L_S$ 155° (Lee, 1986; Christensen, 1988). After the 2001 GDS that occurred earlier in the storm season (from $L_S$ 180° to 270°) the area was blanketed by dust except in a narrow central segment (Szwast, et al., 2006; Cantor, 2007). Most of this dust was then removed in southern Syrtis, essentially over a few days period that occurred 50° of $L_S$ after the end of the GDS at $L_S$ 315°. Cleaning was due to horizontal regional winds as indicated by the progressive formation of dark streaks cutting through the dust, later on resulting in bright streaks in the lee of craters (Sagan, et al., 1973; Szwast, et al., 2006). While further cleaning occurred later on in the eastern part of Syrtis (Cantor, 2007), the western part remained largely covered by dust. Some bright areas not modified during the GDS in the north of Syrtis were cleaned after the GDS at $L_S$ 315° (Szwast, et al., 2006).

We compare in Figure 17 various observations of Syrtis Major obtained from 1971 to 2010. The difference between 1999 and 2002 observations (reduction of Syrtis width due to dust in the west margin) is similar (but opposite) to the difference observed between 1971 and 1978 observations. Observations obtained in 2004 reveal a Syrtis feature consistent with the 2002 observation, i.e. still narrower compared to 1999 with some dust covering the west margin (Figure 17). Most of the area has been observed twice by OMEGA in MY27, at $L_s$ 9-55° (2004) and then at $L_s$ 310-339° (2005): the area is stable over that period with albedo variations within aerosols uncertainties. Small-scaled darkening is observed at two places in MY28 observations obtained prior to the GDS during the storm season (at $L_s$ 133° (2006) and $L_s$ 218-232° (2007)). Elsewhere, most of the dust deposited during the 2001 GDS in the west margin of Syrtis is still there 3 Martian years later. Some observations were acquired over the area during the MY28 GDS in August 2007. In particular, a set of 3 observations obtained between $L_s$ 293° (8° of $L_s$ after GDS maximum) and $L_s$ 300° reveal that dust was removed by horizontal winds during GDS maximum: the dark areas extended westward with remaining bright streaks in the lee of craters. The cleaning of Syrtis Major during MY28 GDS is further revealed by the next observations gathered throughout MY29 (from $L_s$ 16°) and MY30 (to $L_s$ 95°): dark areas covered by dust during 2001 GDS have been recovered in the west and north east areas. The area is now close to its 1999 appearance in the north and 1978 appearance in the west, while being still partly dust covered compared to 1999 in the west. There is no evidence for significant change over MY29 and MY30 considering aerosols uncertainties.

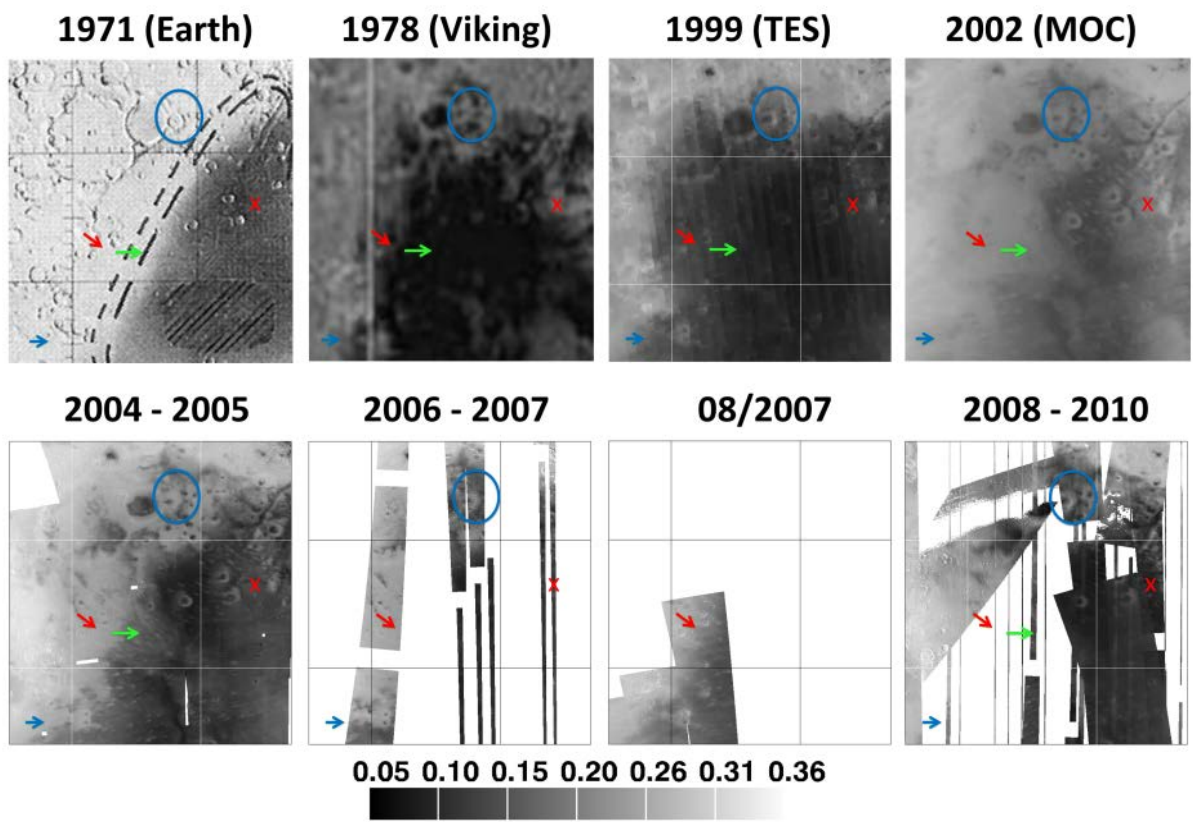

Figure 17: Observations of Syrtis Major from 1971 to 2010. North is on top. Grid corresponds to longitudes 60°E, 70°E and latitudes 10°N, 20°N. Top: previous observations (adapted from (Capen, 1976; Geissler, 2005; Putzig & Mellon, 2007)). Bottom: selected OMEGA observations ("2004-2005": $L_s$ 9° to 55° and 310° to 339° of MY27; "2006-2007": $L_s$ 133° and 218° to 232° of MY28; "08/2007": $L_s$ 293° to 300° of MY28; "2008 – 2010": $L_s$ 16° to 101°, 187° to 246° and 326° of MY29 to 95°of MY30). Color symbols highlight darkening between 2004 and 2010 (blue: darkening prior to MY28 GDS; red and green: darkening during the GDS). Scale bar is indicated for OMEGA and TES albedo only.

To summarize, the evolution of Syrtis Major during the 2007 GDS is significantly different from that observed previously, as no major dust deposits occurred during the storm. On the contrary, the west and north east areas, previously covered by dust during 2001 GDS, were cleaned during the 2007 GDS. Due to the difference in 2001 and 2007 GDS timing, the 2007 GDS decay occurred close to major cleaning events observed in 2001. This could be the main reason for the different behavior. The timing of the 1977 second GDS was similar to the 2007 GDS while persistent brightening was observed in 1977. However, the observed albedo increase was moderate (0.16 versus 0.13) and the story in 1977 was probably different due to the occurrence of two GDS during the same year. Overall, changes are mainly related to GDS but they are not similar from one GDS to the next. Over several decades, the successive action of GDS seems to result in a cyclic evolution at Syrtis Major, the area being alternately narrower or wider.

### 4.4 Southern hemisphere

#### *4.4.1 Overview*

We observe significant albedo changes at most longitudes of the southern hemisphere which is essentially compose of dark or intermediate albedo area. Important differences are detected when we compare pre and post GDS observations (Figure 9). Spectacular albedo changes in the southern hemisphere were previously reported when the comparison interval includes a GDS: between 1972 and 1976 (Veverka, et al., 1977; Zurek & Martin, 1993), after Viking 1977 GDS (Bell, et al., 1999; Geissler, 2005), after the 2001 GDS (Szwast, et al., 2006; Cantor, 2007).

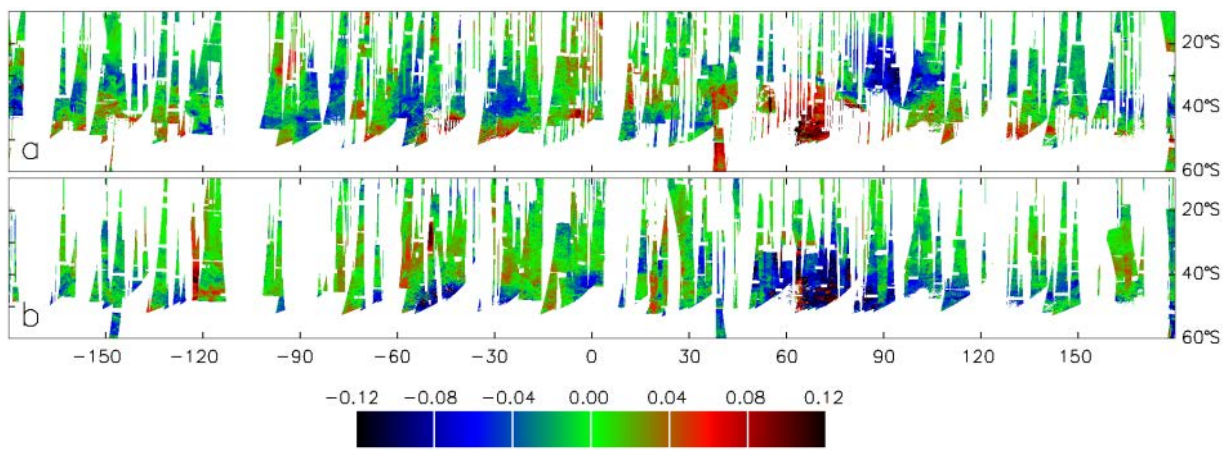

Figure 18: difference albedo maps (southern hemisphere). (a) MY29 clear – MY27 dusty (effect of the MY28 GDS); (b) MY29 dusty – MY29 clear (post-GDS variability over a year). See Table 2 for definition of albedo maps.

Part of the changes introduced by the GDS are temporary as they are compensated by reverse changes during the following storm season (Figure 18b compare to Figure 18a). Seasonal darkening or brightening of the southern latitudes was observed all over the last century, notably about $L_s$ 200° (Capen, 1976; De Mottoni Y Palacios & Dollfus, 1982; Lee, 1986). This is particularly true for MY28 GDS dust deposits which persist over MY29 clear season but are frequently removed during MY29 storm season, similarly to what was observed in South Utopia (Figure 13). On the other hand, dust removals that occurred during the MY28 GDS are primarily not compensated afterward. A similar situation was observed in the southern hemisphere after MY25 GDS (Szwast, et al., 2006): southern hemisphere GDS dust deposits were removed the following year (except in Hesperia), while the large scale darkening in Daedalia was persistent over the next year. As pointed out previously (Szwast, et al., 2006), the strong seasonal activity in the southern hemisphere is linked with the great equatorward extent of the seasonal cap, down to about 45°S (Langevin, et al., 2007), as dust storm activity is frequent near cap edges due to thermal winds (Cantor, et al., 2001).

### 4.4.2 Hesperia

A large-scale darkening occurred in Hesperia, north east of the Hellas basin (Figure 9). A moderately bright (albedo about 0.20 - 0.22) area prior to the MY28 GDS, 1500 x 1000 km wide, gets entirely cleaned after the GDS with albedo reduced to 0.13-0.15. The same area was also subject to change during the MY25 GDS (Szwast, et al., 2006): however, while we observe a darkening of the area, a brightening (from 0.12 to 0.18), persistent the following year, was observed in 2001. We can then conclude that dust deposited during the MY25 (2001) GDS have been preserved during 3 Martian years and was then removed under the action of the MY28 (2007) GDS. No major changes are observed in that area in MY29 and MY30 following observations (i.e., no return of the dust cover).

### 4.4.3 Hellas

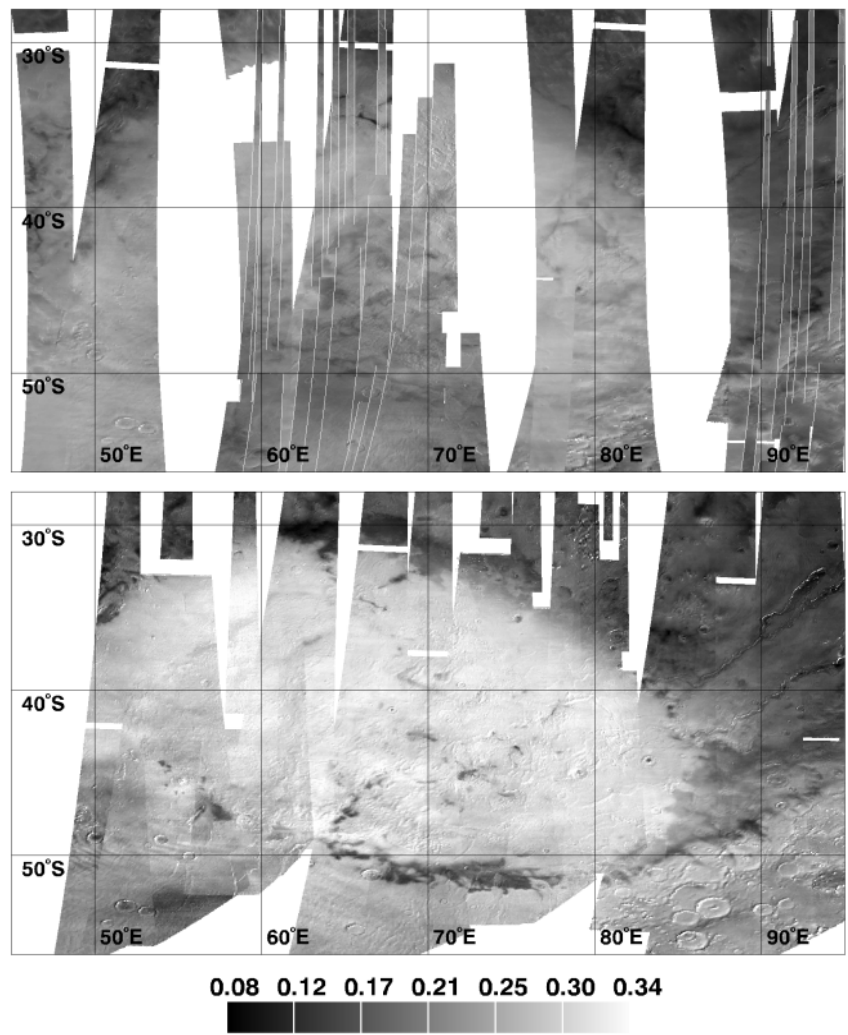

Figure 19: OMEGA observations of Hellas albedo. Top: observations obtained between $L_S$ 252° and $L_S$ 280° in MY27 (white outlined) are superimposed on observations collected after MY29 polar cap retreat ($L_S$ 200-258° of MY29 and $L_S$ 3-10° of MY30): Hellas surface albedo is similar at these epochs (the Hesperia darkening is apparent about 92°E and 35°S). Bottom: observations obtained in MY29 after the MY28 GDS, between $L_S$ 19° and $L_S$ 54°.

The Hellas Basin (Figure 19) is also subject to significant changes with timing related to the fact that Hellas is covered by the seasonal cap in winter. Hellas surface was moderately bright prior to the 2007 GDS (albedo about 0.17-0.25) in most places, with a very few dark patches with lower albedo values (0.11 -0.13). Hellas surface was largely covered by dust during the storm (albedo about 0.3 – 0.32), while some area in the south remained unchanged or became even darker (albedo down to 0.13). An albedo increase up to 0.3 – 0.32 suggests an optically thick layer of dust. Indeed, some of the small scale dark features present prior to the GDS are not discernable anymore after the GDS (Figure 19). Then, this dust is entirely removed once the polar cap retreated at $L_S$ 200°, with albedo absolute values and pattern exactly similar to MY27 observations. Qualitatively, a similar evolution was observed during the 2001 GDS (Szwast, et al., 2006).

# 5  Summary and conclusions

We have developed a method to compute the hemispherical solar albedo of the surface of Mars using OMEGA data. Comparison of measurements obtained by OMEGA and TES in similar conditions reveals a very good agreement between both instruments in terms of absolute calibration (3% difference). Once corrected for atmospheric radiative transfer and surface photometry, OMEGA dark surface albedos are comparable to previous estimate by TES while OMEGA bright surface albedos are on average 17% higher. We have derived a global non polar albedo map (between 60°S and 60°N with 99% filling) at 40 pixels per degree spatial sampling from observations with typical spatial resolution between 1 and 2 km. Observations were obtained from late MY26 to mid MY30 (2004 - 2010). We have also built six partly filled (25% to 75%) seasonal maps over that period. This albedo dataset has been used in companion studies (e.g., (Audouard, et al., 2014)) to provide an accurate estimate of the energy balance at the surface of Mars.

Accounting for the contribution of aerosols is essential for robust interpretations of surface changes between OMEGA observations. Not only does the contribution of aerosols vary with time and place, but it also changes with photometric angles which range is extended in the OMEGA dataset. While automatically removing the full impact of aerosols for the whole dataset remains out of reach, notably due to the lack of simultaneous – in time and space – measurements of the visible optical depth, it has been possible to develop a first order aerosols correction. Uncertainties and bias linked with this correction have been estimated, allowing robust identifications of surface changes.

Surface albedo changes (brightening and darkening) are detected all over the mission. In agreement with previous interpretations, all changes are fully consistent with the deposition or removal of bright dust over darker terrains. Most changes occur in low or intermediate albedo terrains (0.1 – 0.2), while bright surfaces (albedo 0.3), usually considered as covered by an optically thick layer of dust, remain essentially steady. Large-scaled albedo increases up to 0.3 are only observed in the Hellas basin and come with the disappearance of some underlying dark features. In most other brightening cases, the shape of the underlying dark surface can still be perceived. As a consequence, an optically thin layer of bright dust is involved in most albedo changes, micrometers to tens of µm thick only according to earlier work (Christensen, 1988; Cantor, 2007; Kinch, et al., 2007). As noticed previously, this is in agreement with the relatively high thermal inertia of these changing areas (Geissler, 2005). This is also consistent with the lack of associated thermal inertia changes observed in earlier works (Lee, 1986) and in this study: dark, high thermal inertia material is thermally seen through the dust, while dust is thick enough to increase albedo.

The thin dust layer involved in albedo changes masks the underlying mineralogy detected with near-IR spectroscopy. This is illustrated by an impressive albedo change that occurred in the area of Propontis (182°E, 37°N). A dark area (albedo 0.1) several hundreds of km wide gets covered and almost erased by a thin dust layer (albedo increase to 0.25) in 2009. The intermediate resulting albedo and the fact that the dark feature was still perceivable indicate that the dust cover is optically thin. This thin layer of bright dust is however sufficient to mask the strong pyroxenes signatures previously detected. As this dust layer has not been removed by 2014 (Lee, et al., 2014), this example shows that surface mineralogy maps derived from near-IR spectroscopy are only representative of the upper, μm to a tens of μm thick layer of material and can change depending on observation years due to dust redistribution.

Most changes occur during the storm season ($L_S$ 130° – 0°). Changes are not progressive: they appear from one observation to the next. They are thus best explained by dust removal or dust deposition linked with isolated events: dust storms, in agreement with previous studies (Szwast, et al., 2006). Changes frequently occurred near boundaries. This suggests regional dust redistribution due to nearby dust lifting and deposition where thermal contrasts and associated winds are high, in agreement with previous observations and analyzes (Cantor, et al., 2001; Szwast, et al., 2006; Cantor, 2007). Such regional horizontal transport of bright material erasing darker underlying ground and associated with seasonal dust storms have also been observed locally with in-situ exploration (Geissler, et al., 2010). Changes are numerous during the MY28 Global Dust Storm (GDS) and during the following MY29 storm season, while being of lower amplitude during MY27 which postdates by two years the previous GDS. Similarly, strong activity was also reported during MY25 GDS and during the following MY26 storm season (Szwast, et al., 2006; Cantor, 2007).

Regional, optically thin dust deposition is observed in some dark areas during the MY28 GDS. Brightening can be limited in areal extent, such as observed in Alcyonius. This is either consistent with localized dust deposition from a regional dust storm or localized wind preservation from dust-fallout. This dust generally remained stable during the subsequent clear MY29 season and was then partly or totally removed during the storm season of MY29 (e.g., Alcyonius, Utopia, Hellas). The late timing of MY28 GDS, up to $L_S$ 340°, probably prevented more rapid dust removal in some places, compared to the MY25 GDS that ended earlier at $L_S$ 270° with dust removal observed during the remaining MY25 storm season (e.g., south Utopia). Hellas cleaning is specific as Hellas is covered by the south seasonal cap in winter: cleaning is revealed during retreat of the cap and is probably linked with $CO_2$ sublimation or thermal winds and associated dust storms observed around the cap.

No global fall-out blanketing of dust is observed after the MY28 GDS. On the contrary, some dark areas well known for their disappearance after previous GDS, in particular Syrtis Major (Lee, 1986; Christensen, 1988; Szwast, et al., 2006; Cantor, 2007), extended during the maximum activity of MY28 GDS. The dark Propontis feature was identical right after GDS decay compared to pre-GDS observations. This indicates that strong winds must have preserved these dark areas from dust fall-out. Indeed, Syrtis Major cleaning during MY28 GDS comes with the formation of bright streaks in the lee of craters, indicative of horizontal winds cleaning. The difference between MY25 and MY28 GDS impact may be linked with timing differences. The MY25 GDS occurred early in the storm season from L$_S$ 180 to 270°, and cleaning events were observed later on (L$_S$ > 300°). The MY28 GDS occurred later in the storm season between L$_S$ 265° and L$_S$ 340°; thus, late storm season cleaning events have occurred simultaneously with MY28 GDS decay, resulting in a lack of persistent dust cover. It has also been suggested that the MY28 GDS (similarly to the MY25 GDS) was mild compared to earlier GDS, which may contribute to explain the lack of large-scale dust blanketing (P. Geissler, personal communication).

Dust deposited during MY25 GDS in western Syrtis and Hesperia persisted up to the onset of the MY28 GDS, and was partly cleaned (western Syrtis) or totally cleaned (Hesperia) during the MY28 GDS. This indicates that albedo changes observed in a given GDS are also the result of the previous GDS. The lack of significant persistent dust cover in Syrtis Major after MY28 GDS may then also result from the fact that the parts of Syrtis that are the more subject to persistent dust cover were already covered. Not all persistent dust patches deposited during MY25 were removed during MY28 GDS, such as observed locally (tens of km wide areas) in south Utopia.

At a global scale, we do not observe over 2004 - 2010 any evidence for steady processes responsible for regular advance or retreat of albedo frontiers, similarly to (Szwast, et al., 2006) over 1999-2004, which further question that such steady processes where mainly responsible for changes observed 20 years after Viking (Bell, et al., 1999; Geissler, 2005). However, we observe that the successive action of sporadic dust removal or deposition can result in an apparent progressive growth of a given area over decades (e.g., Alcyonius). Such trends do not result from a perfectly regular process as it breaks off due to temporary reverse events.

The evolution of certain areas undergoing sporadic, irregular dust removal or deposition, in particular during GDS, appears somehow cyclic over long periods of several decades (e.g., Syrtis Major, Hyblaeus). Other areas undergoing similar isolated modifying events have not yet returned to their appearance of several decades ago (e.g., Cerberus, Propontis). Our observations do not highlight the existence of large scaled perennial albedo changes that could significantly impact the climate, as hypothesized previously (Fenton, et al., 2007).

Overall, a wide variety of types and timing of changes thus exist on Mars: dust deposited during GDS and removed during the following storm season (e.g., Hellas, Alcyonius), dust removed during GDS (e.g., Hyblaeus), dust deposited or removed during a classical storm season followed or preceded by a long period (several years to decades) of stability (e.g., Propontis, Hyblaeus), areas over which dust is deposited and removed during successive GDS essentially (Syrtis Major, Hesperia). A given region can combine several of these change types (e.g., South Utopia). The timing and duration of some significant changes, in particular the Cerberus brightening, are still unknown as no reverse changes have been observed there so far.

The lack of repeatability over successive years or successive GDS years is consistent with the erratic and unpredictable occurrence of the major dust redistribution processes: global dust storms (Zurek & Martin, 1993; Mulholland, et al., 2013). Over longer timescales of decades, the near cyclic evolution of certain areas that results from the successive action of divergent events is consistent with the persistence of Mars main albedo markings over the last century.

# 6 Acknowledgements:

The authors would like to acknowledge the OMEGA engineering and scientific team for making this work possible, in particular J.P. Bibring and Y. Langevin. We also thank M. J. Wolff, N. Putzig, L. Montabone, B. Gondet, G. Carrozzo, Y. Daydou and F. Poulet for their specific help during this project. We finally thank P. Geissler and an anonymous reviewer for their attentive reading of the manuscript and their helpful comments.

# 7 References

Audouard, J. et al., 2014. Mars surface thermal inertia and heterogeneities from OMEGA/MEX. *Icarus,* Volume 233, pp. 194-213, doi: 10.1016/j.icarus.2014.01.045.

Basu, S., Wilson, J., Richardson, M. & Ingersoll, A., 2006. Simulation of spontaneous and variable global dust storms with the GFDL Mars GCM. *Journal of Geophysical Research,* Volume 111, pp. E09004, doi: 10.1029/2005JE002660.

Baum, W. A., 1974. Earth-based observations of Martian albedo changes. *Icarus,* Volume 22, pp. 363-370, doi : 10.1016/0019-1035(74)90183-3.

Bell, J. F. & Ansty, T. M., 2007. High spectral resolution UV to near-IR observations of Mars using HST/STIS. *Icarus,* 191(2), pp. 581-602, doi: 10.1016/j.icarus.2007.05.019.

Bell, J. F. et al., 1999. Near-Infrared Imaging of Mars from HST: Surface Reflectance, Photometric Properties, and Implications for MOLA Data. *Icarus,* Volume 138, pp. 25-35, doi: 10.1006/icar.1998.6057.

Bellucci, G. et al., 2006. OMEGA/Mars Express: Visual channel performances and data reduction techniques. *Planetary and Space Science,* 54(7), pp. 675-684, doi: 10.1016/j.pss.2006.03.006 .

Bonello, G. et al., 2005. The ground calibration setup of OMEGA and VIRTIS experiments: description and performances. *Planetary and Space Science,* 53(7), pp. 711-728, doi: 10.1016/j.pss.2005.02.002.

Bridges, N. T. et al., 2012. Planet-wide sand motion on Mars. *Geology,* 40(1), pp. 31-34.

Cantor, B. A., 2007. MOC observations of the 2001 Mars planet-encircling dust storm. *Icarus,* 186(1), pp. 60-96, doi: 10.1016/j.icarus.2006.08.019.

Cantor, B. A., James, P. B., Caplinger, M. & Wolff, M. J., 2001. Martian dust storms: 1999 Mars Orbiter Camera observations. *Journal of Geophysical Research,* 106(E10), pp. 23653-23687, doi: 10.1029/2000JE001310.

Cantor, B. A., Kanak, K. M. & Edgett, K. S., 2006. Mars Orbiter Camera observations of Martian dust devils and their tracks (September 1997 to January 2006) and evaluation of theoretical vortex models. *Journal of Geophysical Research,* Volume 111, pp. E12002, doi: 10.1029/2006JE002700.

Capen, C. F., 1976. Martian albedo feature variations with season : data of 1971 and 1973. *Icarus,* Volume 28, pp. 213-230, doi: 10.1016/0019-1035(76)90034-8.

Carrozzo, F. G. et al., 2012. Iron mineralogy of the surface of Mars from the 1 µm band spectral properties. *Journal of Geophysical Research,* Volume 117, pp. E00J17, doi: 10.1029/2012JE004091.

Chaikin, A. L., Maxwell, T. A. & El-baz, F., 1981. Temporal changes in the Cerberus region of Mars: Mariner 9 and Viking comparisons. *Icarus,* Volume 45, pp. 167-178, doi: 10.1016/0019-1035(81)90012-9.

Chassefiere, E., Drossart, P. & Korablev, O., 1995. Post-Phobos model for the altitude and size distribution of dust in the low Martian atmosphere. *Journal of Geophysical Research,* 100(E3), pp. 5525-5539, doi: 10.1029/94JE03363.

Chojnacki, M. et al., 2014. Persistent aeolian activity at Endeavour crater, Meridiani Planum, Mars; New observations from orbit and the surface. *Icarus,* Volume in press.

Christensen, P. R., 1988. Global albedo variations on Mars: implications for active aeolian transport, deposition, and erosion. *Journal of geophysical research,* 93(B7), pp. 7611-7624, doi: 10.1029/JB093iB07p07611.

Christensen, P. R. et al., 2001. Mars Global Surveyor Thermal Emission Spectrometer experiment: Investigation description and surface science results. *Journal of Geophysical Research,* 106(E10), pp. 23823-23872, doi: 10.1029/2000JE001370.

De Mottoni Y Palacios, G. & Dollfus, A., 1982. Surface marking variations of selected areas on Mars. *Astronomy and Astrophysics,* Volume 116, pp. 323-331.

Drube, L. et al., 2010. Magnetic and optical properties of airborne dust and settling rates of dust at the Phoenix landing site. *Journal of Geophysical Research,* 115(B9), pp. E00E23, doi: 10.1029/2009JE003419.

Dundas, C. M., Diniega, S. & McEwen, A. S., 2014. Long-term monitoring of Martian gully formation and evolution with MRO/HiRISE. *Icarus,* p. in press.

Erard, S., 2000. The 1994-1995 apparition of Mars observed from Pic-du-Midi. *Planetary and space Science,* Volume 48, pp. 1271-1287, DOI: 10.1016/S0032-0633(00)00109-4.

Fenton, L. K., Geissler, P. E. & Haberle, R. M., 2007. Global warming and climate forcing by recent albedo changes on Mars. *Nature,* 446(7136), pp. 646-649, doi:10.1038/nature05718.

Fergason, R. L., Christensen, P. R. & Kieffer, H. H., 2006. High-resolution thermal inertia derivation from THEMIS: Thermal model and applications. *Journal of Geophysical Research,* Volume 111, pp. E12004, doi: 10.1029/2006JE002735.

Fernando, J. et al., 2013. Surface reflectance of Mars observed by CRISM/MRO: 2. Estimation of surface photometric properties in Gusev Crater and Meridiani Planum. *Journal of Geophysical Research: Planets,* Volume 118, pp. 534-559, doi: 10.1029/2012JE004194.

Forget, F. et al., 1999. Improved general circulation models of the Martian atmosphere from the surface to above 80 km. *Journal of Geophysical Research,* 104(E10), pp. 24155-24176, DOI: 10.1029/1999JE001025.

Forget, F. et al., 2007. Remote sensing of surface pressure on Mars with the Mars Express/OMEGA spectrometer: 1. Retrieval method. *Journal of Geophysical Research,* 112(E8), pp. E08S15, doi:10.1029/2006JE002871.

Fröhlich, C. & Lean, J., 2004. Solar radiative output and its variability: evidence and mechanisms. *The Astronomy and Astrophysics Review,* 12(4), pp. 273-320, doi: 10.1007/s00159-004-0024-1.

Geissler, P. E., 2005. Three decades of Martian surface changes. *Journal of geophysical research,* Volume 110, pp. E02001, doi: 10.1029/2004JE002345.

Geissler, P. E., 2012. *Persistent surface changes in solis lacus, Mars.* The Woodlands, Texas, U.S.A., s.n., p. 2598.

Geissler, P., Enga, M. & Mukherjee, P., 2013. *Global Monitoring of Martian Surface Albedo Changes from Orbital Observations.* s.l., American Geophysical Union, Fall Meeting 2013, abstract #P41A-1911.

Geissler, P. E. et al., 2010. Gone with the wind: Eolian erasure of the Mars Rover tracks. *Journal of Geophysical Research,* 115(E12), pp. E00F11, doi: 10.1029/2010JE003674.

Goetz, W. et al., 2005. Indication of drier periods on Mars from the chemistry and mineralogy of atmospheric dust. *Nature,* 436(7047), pp. 62-65, doi:10.1038/nature03807.

Greeley, R. et al., 2005. Martian variable features: New insight from the Mars Express Orbiter and the Mars Exploration Rover Spirit. *Journal of Geophysical Research,* 110(E6), pp. E06002, DOI: 10.1029/2005JE002403.

Greeley, R. et al., 2006. Active dust devils in Gusev crater, Mars: Observations from the Mars Exploration Rover Spirit. *Journal of Geophysical Research,* 111(E12), pp. E12S09, DOI: 10.1029/2006JE002743.

Hapke, B., 1993. *Theory of reflectance and emittance spectroscopy.* New York: Cambridge University Press.

James, P. B. et al., 1996. Global imaging of Mars by Hubble space telescope during the 1995 opposition. *Journal of Geophysical Research,* 101(E8), pp. 18883-18890, DOI: 10.1029/96JE01605.

Jouglet, D., 2008. *L'hydratation de la surface de Mars vue par l'imageur spectral OMEGA,* Orsay, France: Université Paris Sud 11.

Jouglet, D. et al., 2009. OMEGA long wavelength channel: Data reduction during non-nominal stages. *Planetary and Space Science,* 57(8-9), pp. 1032--1042, DOI: 10.1016/j.pss.2008.07.025.

Kahre, M. A. et al., 2005. Observing the martian surface albedo pattern: Comparing the AEOS and TES data sets. *Icarus,* 179(1), pp. 55-62, doi: 10.1016/j.icarus.2005.06.011.

Kieffer, H. H., 2013. Thermal model for analysis of Mars infrared mapping. *Journal of Geophysical Research,* Volume 118, pp. 451-470, doi:10.1029/2012JE004164.

Kinch, K. M. et al., 2007. Dust deposition on the Mars Exploration Rover Panoramic Camera (Pancam) calibration targets. *Journal of Geophysical Research,* 112(E6), pp. E06S03, DOI: 10.1029/2006JE002807 .

Landis, G. A. & Jenkins, P. P., 2000. Measurement of the settling rate of atmospheric dust on Mars by the MAE instrument on Mars Pathfinder. *Journal of Geophysical Research,* 105(E1), pp. 1855-1857, DOI: 10.1029/1999JE001029.

Langevin, Y., 2007. *OMEGA Instrumental problems.* Villafrance, Spain, First Mars Express data workshop - HRSC & OMEGA.

Langevin, Y. et al., 2007. Observations of the south seasonal cap of Mars during recession in 2004-2006 by the OMEGA visible/near-infrared imaging spectrometer on board Mars Express. *Journal of Geophysical Research,* Volume 112, pp. E08S12, DOI: 10.1029/2006JE002841.

Lee, S. W., 1986. *Regional sources and sinks of dust on Mars: Viking observations of Cerberus, Solis Planum, and Syrtis Major.* Washington, Symposium on Mars: Evolution of its climate and atmosphere, page 57.

Lee, S. W. & Clancy, R. T., 1990. *The Effects of Atmospheric Dust on Observations of the Surface Albedo of Mars.* Houston, Lunar and Planetary Science Conference, 21, 688.

Lee, S. W., Thomas, P. C. & Cantor, B. A., 2014. *Disappearance of the Propontis Regional Dark Albedo Feature on Mars.* Pasadena, Eight International Conference on Mars, abtract 1426.

Leighton, R. B. et al., 1969. Mariner 6 and 7 television pictures: preliminary analysis. *Science,* 166(3901), pp. 49-67, DOI: 10.1126/science.166.3901.49 .

Lemmon, M. T. et al., 2015. Dust aerosol, clouds, and the atmospheric optical depth record over 5 Mars years of the Mars Exploration Rover mission. *Icarus,* Volume 251, pp. 96-111.

Lemmon, M. T. et al., 2004. Atmospheric Imaging Results from the Mars Exploration Rovers: Spirit and Opportunity. *Science,* 306(5702), pp. 1753-1756, DOI: 10.1126/science.1104474 .

Liu, J., Richardson, M. I. & Wilson, R. J., 2003. An assessment of the global, seasonal, and interannual spacecraft record of Martian climate in the thermal infraredJournal of Geophysical Research. *Journal of Geophysical Research,* 108(E8), pp. 8-1, DOI: 10.1029/2002JE001921.

Madeleine, J.-B.et al., 2012. Aphelion water-ice cloud mapping and property retrieval using the OMEGA imaging spectrometer onboard Mars Express. *Journal of Geophysical Research,* Volume 117, pp. E00J07, doi:10.1029/2011JE003940.

Malin, M. C. et al., 2006. Present-day impact cratering rate and contemporary gully activity on Mars. *Science,* Volume 314, pp. 1573-1577, DOI: 10.1126/science.1135156 .

McCord, T. B. et al., 2007. Mars Express High Resolution Stereo Camera spectrophotometric data: Characteristics and science analysis. *Journal of Geophysical Research,* 112(E6), pp. E06004, DOI: 10.1029/2006JE002769 .

McEwen, A. S. et al., 2011. Seasonal flows on warm martian slopes. *Science,* Volume 333, pp. 740-743, DOI: 10.1126/science.1204816 .

Mellon, M. T., Jakosky, B. M., Kieffer, H. H. & Christensen, P. R., 2000. High-resolution thermal inertia mapping from the Mars Global Surveyor Thermal Emission Spectrometer. *Icarus,* Volume 148, pp. 437-455, doi:10.1006/icar.2000.6503.

Montabone, L. et al., 2015. Eight-year Climatology of Dust Optical Depth on Mars. *Icarus,* Volume 251, pp. 65-95.

Montmessin, F. et al., 2007. On the origin of perennial water ice at the south pole of Mars: A precession-controlled mechanism?. *Journal of Geophysical Research,* 112(E8), pp. E08S17, DOI: 10.1029/2007JE002902.

Mulholland, D. P., Read, P. L. & Lewis, S. R., 2013. Simulating the interannual variability of major dust storms on Mars using variable lifting thresholds. *Icarus,* 223(1), pp. 344-358, DOI: 10.1016/j.icarus.2012.12.003.

Ody, A. et al., 2012. Global maps of anhydrous minerals at the surface of Mars from OMEGA/MEx. *Journal of Geophysical Research,* Volume 117, pp. E00J14, doi:10.1029/2012JE004117.

Paige, D. A. & Keegan, K. D., 1994. Thermal and albedo mapping of the polar regions of Mars using Viking thermal mapper observations: 2. South polar region. *Journal of Geophysical Research,* 99(E12), pp. 25993-26013, DOI: 10.1029/93JE03429.

Pleskot, L. K. & Miner, E. D., 1981. Time variability of martian bolometric albedo. *Icarus,* Volume 45, pp. 179-201, DOI: 10.1016/0019-1035(81)90013-0.

Pollack, J. B. & Sagan, C., 1967. Secular changes and dark-area regeneration on Mars. *Icarus,* Volume 6, pp. 434-439, DOI: 10.1016/0019-1035(67)90036-X.

Poulet, F. et al., 2007. Martian surface mineralogy from Observatoire pour la Minéralogie, l'Eau, les Glaces et l'Activité on board the Mars Express spacecraft (OMEGA/MEx): Global mineral maps. *Journal of Geophysical Research,* 112(E8), pp. E08S02, DOI: 10.1029/2006JE002840.

Putzig, N. E. & Mellon, M. T., 2007. Apparent thermal inertia and the surface heterogeneity of Mars. *Icarus,* Volume 191, pp. 68-94, DOI: 10.1016/j.icarus.2007.05.013.

Putzig, N. E., Mellon, M. T., Kretke, K. A. & Arvidson, R. E., 2005. Global thermal inertia and surface properties of Mars from the MGS mapping mission. *Icarus,* Volume 173, pp. 325-341, DOI: 10.1016/j.icarus.2004.08.017.

Rice, M. S. et al., 2011. Temporal observations of bright soil exposures at Gusev crater, Mars. *Journal of Geophysical Research,* Volume 116, pp. E00F14, DOI: 10.1029/2010JE003683.

Sagan, C. et al., 1973. Variable Features on Mars, 2, Mariner 9 Global Results. *Journal of Geophysical Research,* 78(20), pp. 4163-4196, DOI: 10.1029/JB078i020p04163.

Sagan, C. et al., 1972. Variable Features on Mars: Preliminary Mariner 9 Television Results. *Icarus,* Volume 17, pp. 346-372, DOI: 10.1016/0019-1035(72)90005-X.

Schorghofer, N. & King, C. M., 2011. Sporadic formation of slope streaks on Mars. *Icarus,* Volume 216, pp. 159-168, DOI: 10.1016/j.icarus.2011.08.028.

Singer, R. B. & Roush, T. L., 1983. *Spectral Reflectance Properties of Particulate Weathered Coatings on Rocks: Laboratory Modeling and Applicability to Mars.* Houston, LUNAR AND PLANETARY SCIENCE XIV, 708-709.

Smith, D. E. et al., 1999. The Global Topography of Mars and Implications for Surface Evolution. *Science,* 284(5419), pp. 1495-1503, DOI: 10.1126/science.284.5419.1495 .

Smith, M. D., 2004. Interannual variability in TES atmospheric observations of Mars during 1999-2003. *Icarus,* 167(1), pp. 148-165, DOI: 10.1016/j.icarus.2003.09.010.

Smith, M. D. et al., 2006. One Martian year of atmospheric observations using MER Mini-TES. *Journal of Geophysical Research,* 111(E12), pp. E12S13, DOI: 10.1029/2006JE002770.

Sullivan, R. et al., 2008. Wind-driven particle mobility on Mars: Insights from Mars Exploration Rover observations at ``El Dorado'' and surroundings at Gusev Crater. *Journal of Geophysical Research,* 113(E6), pp. E06S07, DOI: 10.1029/2008JE003101.

Sullivan, R. et al., 2001. Mass movement slope streaks imaged by the Mars Orbiter Camera. *Journal of Geophysical Research,* 106(E10), pp. 23607-23634, DOI: 10.1029/2000JE001296.

Szwast, M. A., Richardson, M. I. & Vasavada, A. R., 2006. Surface dust redistribution on Mars as observed by the Mars Global Surveyor and Viking orbiters. *Journal of Geophysical Research,* 111(E11), pp. E11008, doi:10.1029/2005JE002485.

Tomasko, M. G. et al., 1999. Properties of dust in the Martian atmosphere from the Imager on Mars Pathfinder. *Journal of Geophysical Research,* 104(E4), pp. 8987-9008, DOI: 10.1029/1998JE900016.

Treiman, A. H., 2003. Geologic settings of Martian gullies: Implications for their origins. *Journal of Geophysical Research,* Volume E4, pp. 8031, DOI: 10.1029/2002JE001900.

Vaughan, A. F. et al., 2010. Pancam and Microscopic Imager observations of dust on the Spirit Rover: Cleaning events, spectral properties, and aggregates. *International Journal of Mars Science and Exploration,* Volume 4, pp. 129-145, doi:10.1555/mars.2010.0005.

Veverka, J., Thomas, P. & Greeley, R., 1977. A study of variable features on Mars during the Viking primary mission. *Journal of geophysical research,* 82(28), pp. 4167-4187, DOI: 10.1029/JS082i028p04167.

Vincendon, M., 2013. Mars surface phase function constrained by orbital observations. *Planetary and Space Science,* Volume 76, pp. 87-95, DOI: 10.1016/j.pss.2012.12.005.

Vincendon, M. et al., 2013. *Mars Albedo Changes During 2004-2010.* The Woodlands, Texas, 44th Lunar and Planetary Science Conference, 2221.

Vincendon, M. et al., 2007. Recovery of surface reflectance spectra and evaluation of the optical depth of aerosols in the near-IR using a Monte Carlo approach: Application to the OMEGA observations of high-latitude regions of Mars. *Journal of Geophysical Research,* 112(E8), pp. E08S13, doi: 10.1029/2006JE002845.

Vincendon, M. et al., 2008. Dust aerosols above the south polar cap of Mars as seen by OMEGA. *Icarus,* 196(2), pp. 488-505, 10.1016/j.icarus.2007.11.034.

Vincendon, M. et al., 2009. Yearly and seasonal variations of low albedo surfaces on Mars in the OMEGA/MEx dataset: Constraints on aerosols properties and dust deposits. *Icarus,* 200(2), pp. 395-405, doi:10.1016/j.icarus.2008.12.012.

Wang, H. & Richardson, M. I., 2015. The Origin, Evolution, and trajectory of Large Dust Storms on Mars during Mars Year 24-30 (1999-2011). *Icarus,* Volume 251, pp. 112-127 (doi: 10.1016/j.icarus.2013.10.033).

Wolff, M. J. & Clancy, R. T., 2003. Constraints on the size of Martian aerosols from Thermal Emission Spectrometer observations. *Journal of geophysical research,* 108(E9), pp. 5097, DOI: 10.1029/2003JE002057.

Wolff, M. J. et al., 2009. Wavelength dependence of dust aerosol single scattering albedo as observed by the Compact Reconnaissance Imaging Spectrometer. *Journal of Geophysical Research,* 114(E9), pp. E00D04, DOI: 10.1029/2009JE003350.

Wolff, M. J. et al., 2010. Ultraviolet dust aerosol properties as observed by MARCI. *Icarus,* 208(1), pp. 143-155, DOI: 10.1016/j.icarus.2010.01.010.

Yen, A. S. et al., 2005. An integrated view of the chemistry and mineralogy of martian soils. *Nature,* 436(7047), pp. 49-54, doi:10.1038/nature03637.

Zurek, R. W. & Martin, L. J., 1993. Interannual Variability of Planet-Encircling Dust Storms on Mars. *Journal of Geophysical REsearch,* 98(E2), pp. 3247-3259, DOI: 10.1029/92JE02936.